\documentclass{jfm}

\usepackage{amsmath}
\usepackage{amsfonts}
\usepackage{amssymb}
\usepackage{graphicx}
\usepackage{gensymb}
\usepackage{caption}
\usepackage{subcaption}
\usepackage{enumitem}
\usepackage{mathtools}
\usepackage{natbib}
\usepackage{color}

\usepackage[capitalise]{cleveref}

\newcommand{\pder}[2] {\frac{\partial #1}{\partial #2}}

\newcommand{\aderxy}[2]{u#2\pder{#1}{x}+v#2\pder{#1}{y}}

\newcommand{\tw}[1] {\tilde{#1}}

\DeclareMathOperator{\sgn} {sgn}

\newcommand{\Ro}{{Ro}}

\newcommand{\dint}{\,\textrm{d}}
\newcommand{\rmi}{{\rm i}}

\DeclareMathOperator{\J}{J}

\title{The propagation and decay of a coastal vortex on a shelf}
\author{Matthew N. Crowe and Edward R. Johnson}
\date{}

\shorttitle{Propagation and decay of a shelf vortex}
\shortauthor{M. N. Crowe \& E. R. Johnson}
\affiliation{Department of Mathematics, University College London, London, WC1E 6BT, UK}
\date{}

\begin{document}

\maketitle

\begin{abstract}
A coastal eddy is modelled as a barotropic vortex propagating along a coastal shelf. If the vortex speed matches the phase speed of any coastal trapped shelf wave modes, a shelf wave wake is generated leading to a flux of energy from the vortex into the wave field. Using a simply shelf geometry, we determine analytic expressions for the wave wake and the leading order flux of wave energy. By considering the balance of energy between the vortex and wave field, this energy flux is then used to make analytic predictions for the evolution of the vortex speed and radius under the assumption that the vortex structure remains self similar. These predictions are examined in the asymptotic limit of small rotation rate and shelf slope and tested against numerical simulations.

If the vortex speed does not match the phase speed of any shelf wave, steady vortex solutions are expected to exist. We present a numerical approach for finding these nonlinear solutions and examine the parameter dependence of their structure.
\end{abstract}


\vspace{0.3cm}
\hspace{0.2cm}\rule{12.3cm}{0.4pt}

\section{Introduction}

The interaction of interior ocean flows with coastal boundaries is a complicated multi-scale problem with important implications for the dissipation of mesoscale energy and the generation of potential vorticity \citep{DEWARETAL,DEREMBLEETAL}. The dissipation of energy by coastal boundaries may be an important component of the ocean energy budget and hence these boundary processes may influence the global ocean circulation and long-term variability \citep{PENDUFFETAT}. Vortices and coastal trapped waves are important components of this flow \citep{IsernFontanetetal2006,DEWARHOGG,HOGGETAL,DEREMBLEETAL,crowe_johnson_2020} and often occur on scales which are not well resolved by global ocean models. Therefore, an understanding of the energetic consequences of these processes is required to accurately parametrise their effects in global ocean models.

Shelf waves are a form of coastal trapped topographic wave in which disturbances propagate along a coastal boundary due to the combined effects of the Coriolis force and offshore depth variations \citep{LeBlondM78,JOHNSON2011}. These waves are dispersive and travel with the coastline to the right (left) in the Northern (Southern) hemisphere. Unlike Kelvin waves, shelf waves have a modal structure in the offshore direction and can exist in barotropic systems with no change in surface elevation; this allows a full spectrum of shelf waves to be captured by simple shallow water models \citep{johnson89}.

Moving bodies have long been known to generate wave-fields when travelling in fluids which support wave-like solutions and the wave generation by solid bodies has been extensively studied, both experimentally \citep{LONG53,Machioneetal2018} and analytically \citep{Fraenkel56,lighthill_1967,bretherton_1967}. It is expected that travelling vortices would similarly generate waves, however, since the only source of energy for the wave-field is the kinetic energy of the vortex, the formation of a wave-field would lead to a loss of vortex energy and hence a decay of the vortex. This leads to a feedback mechanism where the generated modes depend on the properties of the vortex and the vortex decay depends on the wave energy flux. \citet{FLIERLHAINES94} used an adjoint method to examine the decay of a modon on a beta plane. They found that as the vortex decayed, mass was ejected from the rear. Therefore, unlike energy, momentum and enstrophy were not conserved between the vortex and wave-field. The value of the maximum vorticity was argued to be a second conserved quantity and used to make analytical predictions for the decay of the modon speed and radius. \citet{JohnsonCrowe20} and \citet{Croweetal20} estimated the decay of the Lamb-Chaplygin dipole \citep{MeleshkoH94} and Hill's vortex \citep{HILL1894} in rotating and stratified flows by calculating the work done by the leading order wave drag and equating this to the loss of vortex energy. Conservation of maximum vorticity was again used to close the system and shown to be valid using numerical simulations.

Here we consider a simple analytical model of a moving vortex on the boundary of a coastal shelf with the aim of determining the long term evolution. Our vortex is taken to consist of a near semi-circular region of vorticity centred on the boundary. Using the method of images, this may be modelled as a dipolar vortex with the dipole strength determined by the vortex speed and radius. We begin in \cref{sec:setup} by presenting the model and describing the exponential shelf profile used throughout. In \cref{sec:vortex_decay} we consider the generation of shelf waves by a moving vortices. As expected, we observe that a wave-field will only be generated if the vortex speed matches the phase speed of any shelf waves and hence vortices moving faster than, or in the opposite direction to, every shelf wave mode will not generate a wave wake. The generation of these waves leads to a flux of energy from the vortex to the wave-field resulting in a decay of the vortex. We use a simple energy balance to estimate this decay and present analytical results for the case of asymptotically small rotation rate and shelf slope. Our predictions are tested against numerical simulations in \cref{sec:num_sim}. In \cref{sec:nonlin_steady}, we examine the case where the vortex does not generate waves. We expect steady vortex solutions to exist and present a numerical approach for finding these fully nonlinear solutions. Finally, in \cref{sec:diss} we discuss our results and the limitations of our model.

\section{Setup}
\label{sec:setup}

Our starting point is the two-dimensional rotating shallow water equations under the rigid lid assumption. Let $Ox'y$ be Cartesian coordinates, fixed in the topography, where $x'$ describes the distance along a straight coastline and $y$ describes the distance in the offshore direction. We consider a near-semicircular vortex moving along the coastline with speed $U(t)$ and introduce coordinates following the vortex centre \citep{JohnsonCrowe20} by defining
\begin{equation}
x = x'-\int_0^t U(t') \dint t',
\end{equation}
and working in the coordinate system $Oxy$. All quantities are specified by their value in the frame of the topography unless stated otherwise. We thus simply work with topographic frame variables expressed as functions of  the moving coordinates $(x,y)$ rather than working in variables relative to a frame translating with speed $U(t)$. The advantage of this formulation is that fictitious forces, that would arise from treating quantities relative to the accelerating vortex frame,  are absent. Throughout, whenever considering vortices which are semi-circular, or asymptotically close to semi-circular, we will denote the radius by $a(t)$.

\begin{figure}
	\centering
	\begin{subfigure}[b]{0.8\textwidth}
	\centering
	\includegraphics[trim={0 0 0 0},clip,width=\textwidth]{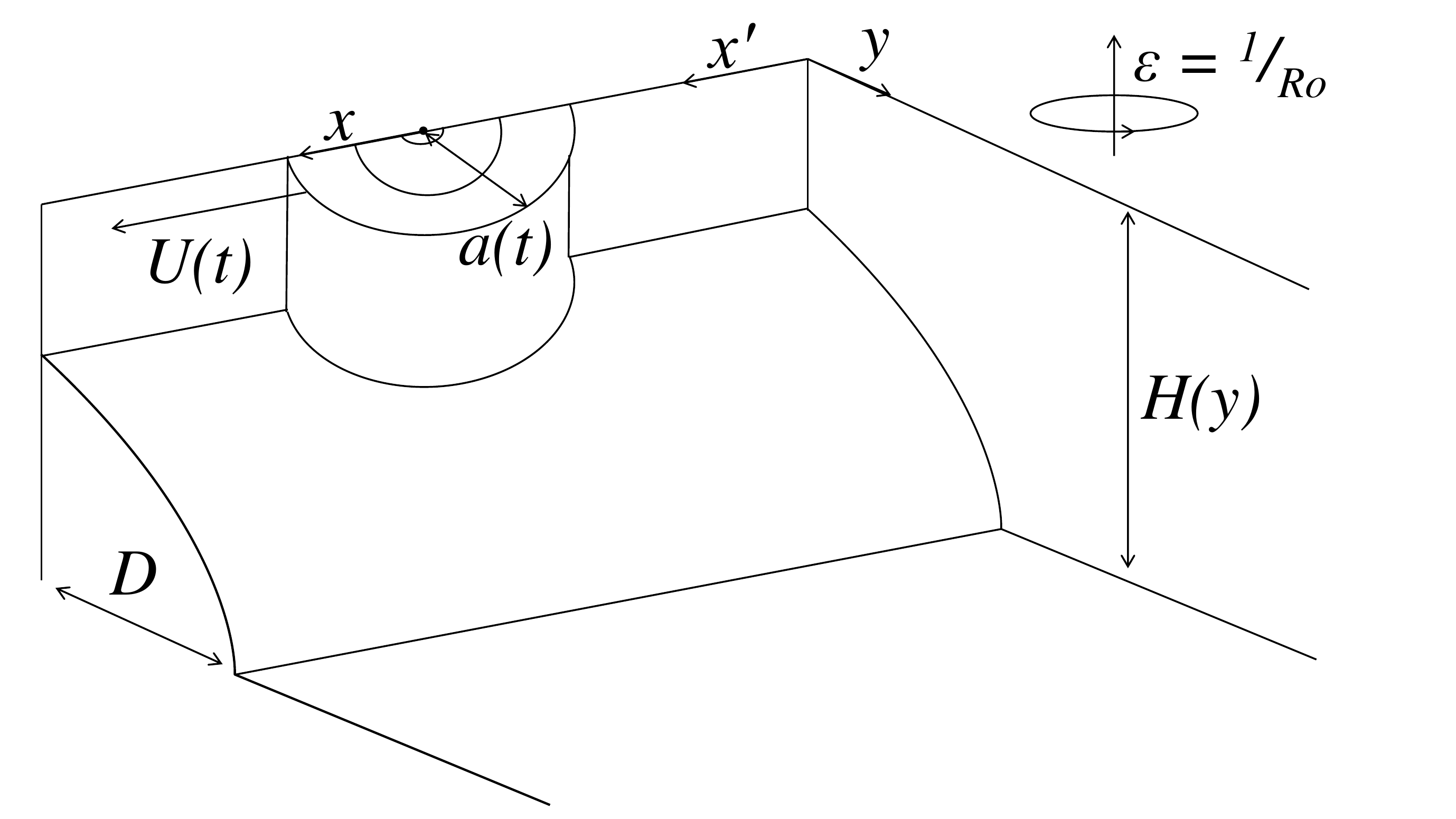}
	\end{subfigure}
	\caption{Our non-dimensional setup showing a vortex of radius $a(t)$ moving along a coastal boundary with speed $U(t)$. The layer depth, $H(y)$, is shown as a shelf region of increasing depth of width $D$ matched to a constant depth ocean. The system is rotating with inverse Rossby number of $\epsilon$.}
    \label{fig:fig1}
\end{figure}

The non-dimensional equations in terms of $(x,y)$ governing the horizontal velocity components $(u,v)$ relative to the topography are thus
\begin{subequations}
\label{eq:gov_eq}
\begin{alignat}{3}
\label{eq:gov_eq_a}
\pder{u}{t}-U\pder{u}{x}+\aderxy{u}{} - \epsilon v &=&\, -\pder{p}{x},\\
\pder{v}{t}-U\pder{v}{x}+\aderxy{v}{} + \epsilon u &=&\, -\pder{p}{y},\\
\pder{}{x}\left(u H\right) + \pder{}{y}\left(v H \right) &=& 0,
\end{alignat}
\end{subequations}
where $\epsilon = 1/\Ro$ is the inverse Rossby number and $H$ is the layer depth. We take $H$ to represent a shelf with the depth varying only in the offshore, $y$, direction so $H = H(y)$. Our setup is shown in \cref{fig:fig1} for a vortex of radius $a(t)$ and a depth profile, $H(y)$, consisting of a shelf of width $D$ joined to a region of constant depth.

\cref{eq:gov_eq} can be combined to give a single evolution equation for the potential vorticity
\begin{equation}
\label{eq:QG}
\left(\pder{}{t}-U\pder{}{x}\right)q + \frac{1}{H}J[\psi,q] = 0,
\end{equation}
where the velocity can be expressed using a volume flux streamfunction
\begin{equation}
(u,v) = \frac{1}{H}\left(-\pder{\psi}{y},\pder{\psi}{x}\right),
\end{equation}
the vorticity is given by
\begin{equation}
\zeta = \pder{v}{x}-\pder{u}{y} = \frac{1}{H}\pder{^2\psi}{x^2}+\pder{}{y}\left[\frac{1}{H}\pder{\psi}{y}\right],
\end{equation}
and $q$ denotes the potential vorticity (PV) in the layer
\begin{equation}
q = \frac{\zeta+\epsilon}{H}.
\end{equation}
We impose a wall at $y = 0$ with the boundary condition $v(x,0) = 0$ or equivalently $\psi(x,0) = 0$. Far from the wall, disturbances are assumed to decay so $(u,v) \to 0$ as $y \to \infty$.

Throughout this study we will consider a shelf profile for $H$ consisting of a shelf region with exponentially increasing depth matched to a constant depth ocean. This profile is chosen for simplicity of calculation however \citet{Huthnance74} and \citet{GillSchumann74} showed that both the form of the dispersion relation remains unchanged and the inner product exists for general topography so similar results are expected to hold for arbitrary $H(y)$. Where possible, general results will also be given in terms of the arbitrary profile $H = H(y)$. Our shelf profile is given by
\begin{equation}
\label{eq:H_def}
H(y) = \begin{cases}
\exp[\beta y], & y \leq D,\\
\exp[\beta D], & y \geq D,
\end{cases}
\end{equation}
where $D$ describes the shelf width and $\beta$ describes the shelf slope.

We take our vortex solution to be dipolar and centred at $(x,y) = (0,0)$ in our vortex-following coordinates. This corresponds to a single vortex in $y>0$ moving by the image effect. The vortex boundary is taken to be close to semi-circular with both the vorticity and streamfunction being continuous across this boundary. In the far field, the vortex appears as an irrotational source doublet (or equivalently, vortex doublet) of strength $\mu$ directed in the positive $x$ direction. Therefore, far from the vortex we have
\begin{equation}
\label{eq:dipole_bc}
\psi = \frac{\mu}{2} \delta(x) \quad \textrm{at} \quad y = 0^+.
\end{equation}
In the limit of $\beta \to 0$ (with $\epsilon$ such that $\beta\epsilon \to 0$), our vortex solution is given by the classical Lamb-Chaplygin dipole \citep{MeleshkoH94} and hence
\begin{equation}
\label{eq:LCD}
\psi= \begin{cases}
-Uy + \frac{2 \J_1 (Kr)y}{\J_0 (Ka) Kr}, \qquad &r<a \\
-Ua^2y/r^2,  &r>a,
\end{cases}
\end{equation}
for $r^2 = x^2+y^2$. Here $a$ is the semi-circular vortex radius, $\J_0$ and $\J_1$ are Bessel functions of the first kind and $K = j_1/a$ where $j_1 \approx 3.8317$ is the first non-zero root of $\J_1$. In this case the dipole strength may be calculated as $\mu = 2\pi Ua^2$. A numerical approach for finding these vortex solutions for a general depth profile, $H$, and rotation rate, $\epsilon$, is given in \cref{sec:nonlin_steady}.

\section{Vortex decay}
\label{sec:vortex_decay}

This shelf system admits shelf wave solutions so if the vortex is travelling in the same direction as the phase speed of these waves (requiring $\epsilon U > 0$), it may generate a wave field. These waves will remove energy from the vortex resulting in vortex decay \citep{FLIERLHAINES94,JohnsonCrowe20,Croweetal20}. Here we determine conditions for the existence of a wave-field and hence determine the parameter values for which a vortex will decay. We then use an asymptotic approach to derive an equation for the decay rate of a vortex under the assumption of small rotation rate and shallow slope.

\subsection{The topographic wave field}

We begin by determining the linear topographic wave solutions admitted by the system in the absence of a vortex. Our wave equation is obtained by linearising \cref{eq:QG} and setting $U = 0$ to get
\begin{equation}
\pder{}{t}\left[ \nabla^2 \psi - \frac{H_y}{H}\pder{\psi}{y}\right] - \frac{\epsilon H_y}{H}\pder{\psi}{x} = 0.
\end{equation}
We now take $H$ to be our shelf profile from \cref{eq:H_def} and assume wavelike solutions of the form $\psi = \sqrt{H}\tw{\phi}(y)\exp(\rmi \omega t - \rmi k x)$ to obtain
\begin{equation}
\left[\pder{^2}{y^2} - k^2 \right] \tw{\phi} = \begin{cases}
-\kappa^2\tw{\phi} & y \leq D,\\
0 & y \geq D,
\end{cases}
\end{equation}
where
\begin{equation}
\label{eq:kappa_def}
\kappa^2(k,\omega) = \frac{\epsilon\beta k}{\omega}-\frac{\beta^2}{4}.
\end{equation}
We impose boundary conditions of $\tw{\phi} = 0$ on $y = 0$ and $\tw{\phi} \to 0$ as $y\to\infty$. Two boundary conditions are also required at $y = D$ so we take the velocity, $(u,v)$, to be continuous here giving that $\tw{\phi}$ is continuous across $y = D$ and
\begin{equation}
\label{eq:u_bc}
\left[ \frac{H_y}{2H} \tw{\phi} + \tw{\phi}_y \right]_{D^-}^{D^+} = 0.
\end{equation}

For $y > D$ our solution is of the form
\begin{equation}
\tw{\phi} = C_1(k) \exp\left(-|k|y\right),
\end{equation}
for some $C_1$ hence, using \cref{eq:u_bc}, we can impose the boundary condition
\begin{equation}
\label{eq:D_bc}
\tw{\phi}_y + \left(\frac{\beta}{2}+|k|\right)\tw{\phi} = 0 \quad \textrm{on} \quad y = D,
\end{equation}
and only consider the shelf region $y \in [0,D]$. In the shelf region we have solution given by
\begin{equation}
\tw{\phi} = C_2(k) \sin \left[\sqrt{\kappa^2-k^2} y\right],
\end{equation}
where the square root term may be complex. Using the boundary condition at $y = D$, \cref{eq:D_bc}, we obtain the dispersion relation
\begin{equation}
\label{eq:disp_rel}
\tan\left[\sqrt{\kappa^2-k^2}D\right] = -\frac{\sqrt{\kappa^2-k^2}}{|k|+\beta/2},
\end{equation}
with solutions describing a countably infinite set of modes with differing wave number and offshore structure. To proceed we define
\begin{equation}
\label{eq:l_def}
l = \sqrt{\kappa^2-k^2},
\end{equation}
so $l$ can be thought of as the offshore wavenumber discretised by the shelf boundary at $y = D$. We can now solve \cref{eq:disp_rel} numerically for $l(k)$ for each mode and plot the frequency, $\omega$, and phase speed, $c_p = \omega/k$, by combining \cref{eq:kappa_def} and \cref{eq:l_def} to get
\begin{equation}
\label{eq:omega_c_p}
\omega = \frac{\epsilon\beta k}{k^2+l^2+\beta^2/4} \quad \textrm{and} \quad c_p = \frac{\epsilon\beta}{k^2+l^2+\beta^2/4}.
\end{equation}
For $\epsilon > 0$, the phase speed of the waves is positive for all wavenumbers and conversely for $\epsilon < 0$, the topographic wave phase speed is always negative. Additionally, the frequency is odd in $k$, so we only need to consider waves with $k \geq 0$. From \cref{eq:disp_rel}, we note that for a given mode, the offshore wavenumber, $l(k)$, is an increasing function of $k$ and lies within the interval
\begin{equation}
\label{eq:l_bound}
l \in \left( \left[n-\frac{1}{2}\right]\!\frac{\pi}{D}, \frac{n\pi}{D} \right),
\end{equation}
for mode number $n = \{1,2,3,\dots\}$ with
\begin{equation}
l(0) \,\,\, \textrm{satisfying}\,\,\tan{Dl} = -\frac{2l}{\beta} \quad\textrm{and} \quad l(k) \to \frac{n\pi}{D} \quad \textrm{as} \quad k\to \infty.
\end{equation}
Since $\omega = c_p k$, the group velocity is given by
\begin{equation}
c_g = \pder{\omega}{k} = c_p + k\pder{c_p}{k},
\end{equation}
where the second term may be shown to be negative for $k \neq 0$ hence $c_g < c_p$ for $k \neq 0$.

\begin{figure}
	\centering
	\begin{subfigure}[b]{0.49\textwidth}
	\centering
	\includegraphics[trim={0 0 0 0},clip,width=\textwidth]{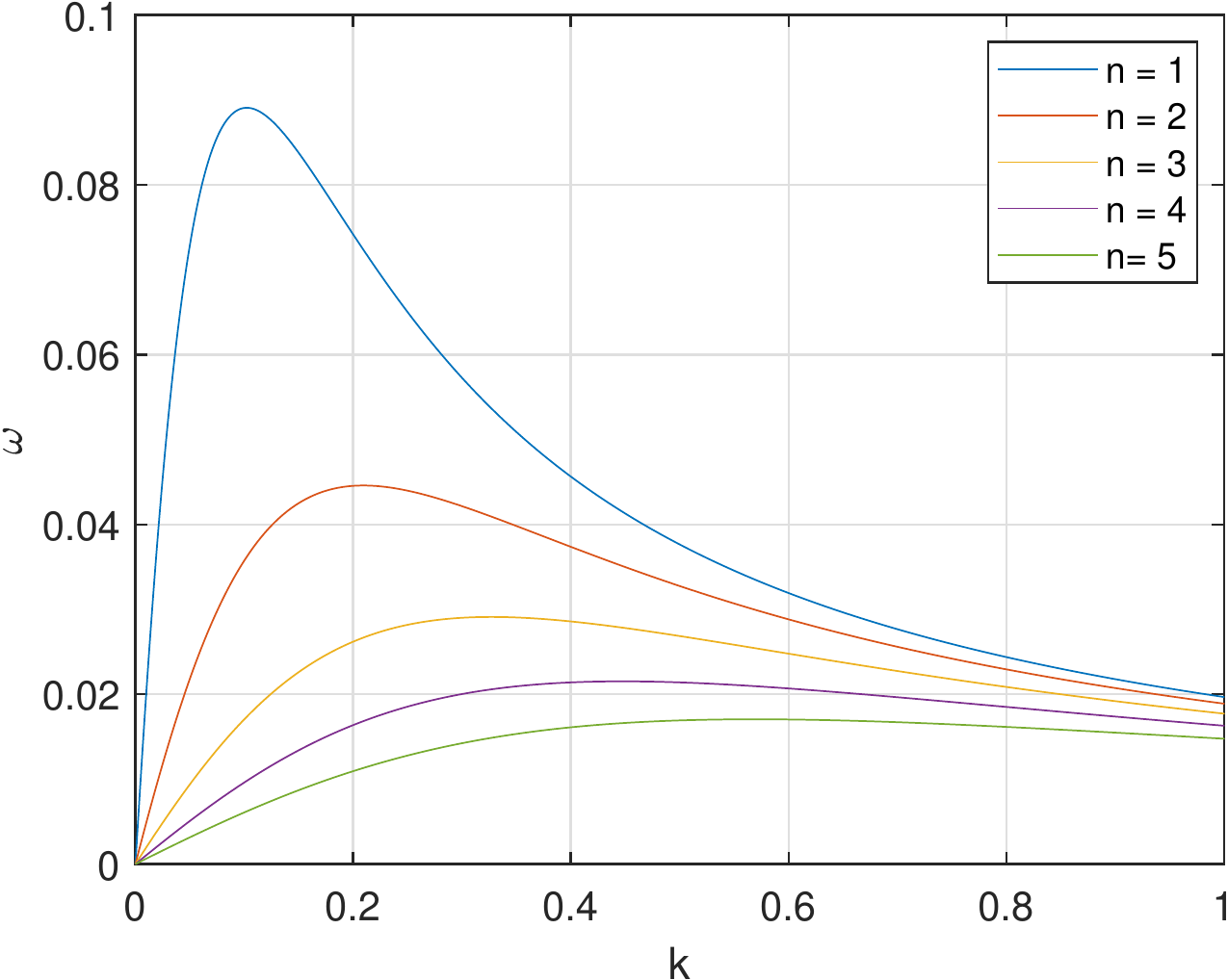}
	\caption{}
	\end{subfigure}
	\begin{subfigure}[b]{0.49\textwidth}
	\centering
	\includegraphics[trim={0 0 0 0},clip,width=\textwidth]{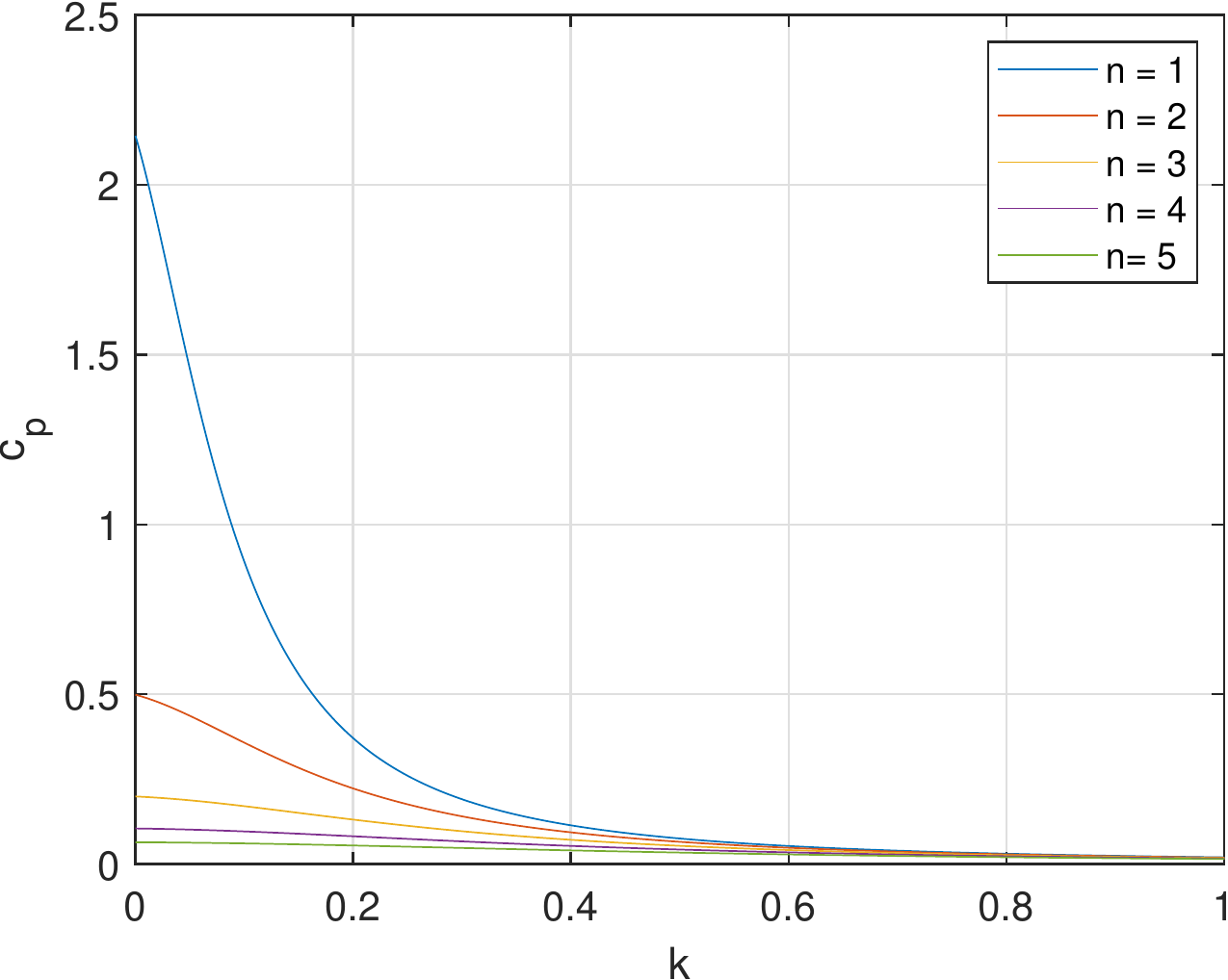}
	\caption{}
	\end{subfigure}
	\caption{Plots of the frequency, $\omega$, (a) and phase speed, $c_p$, (b) for the first five modes with $\epsilon = 0.2$, $\beta = 0.1$ and $D = 25.6$.}
    \label{fig:fig2}
\end{figure}

\cref{fig:fig2} shows the frequency and phase speed for the first five modes with $\epsilon = 0.2$, $\beta = 0.1$ and $D = 25.6$. These curves are consistent with classical results for topographic Rossby waves. For a vortex to generate a wave field, the speed of the vortex must match the phase speed of one or more waves hence a vortex cannot generate any waves if it moves faster than the fastest mode or moves in the opposite direction to the topographic waves. Therefore, for our choice of topography, a vortex moving with speed $U$ will only generate waves if
\begin{equation}
\label{eq:cond_wave}
0 < \epsilon U < \frac{\epsilon^2 \beta}{l_1^2(0) + \beta^2/4},
\end{equation}
where $l_1(0)$ is the smallest solution to $2l+\beta \tan Dl = 0$ and corresponds to the offshore wavenumber of the lowest mode for $k = 0$. Note that $\pi/(2D) < l_1(0) < \pi/D$ and in the case of $\beta \to 0$ we have $l_1(0) \to \pi/(2D)$. The condition in \cref{eq:cond_wave} has been multiplied through by $\epsilon$ to ensure it holds for both positive and negative $\epsilon$. We note that $c_g \leq c_p$ for all modes hence for a vortex moving with speed $U = c_p$, energy will be emitted from the rear of the vortex. This radiation condition will be required later.

If the vortex does not generate waves we expect that steady vortex solutions will exist, these solutions can be found using the method outlined in \cref{sec:nonlin_steady}.

\subsection{Waves generated by a moving vortex}

We now determine the amplitude of the vortex generated wavefield. Working in coordinates following the vortex ($U\neq 0$) and looking for steady wave solutions we obtain the linearised wave equation
\begin{equation}
-U\pder{}{x}\left[\frac{\zeta+\epsilon}{H}\right]-\frac{\epsilon H_y}{H^3}\pder{\psi}{x} = 0.
\end{equation}
Substituting for $\psi = \sqrt{H} \phi$ gives
\begin{equation}
\nabla^2 \phi = \begin{cases}
\left(\frac{\beta^2}{4} - \frac{\epsilon\beta}{U}\right)\phi & y \leq D,\\
0 & y \geq D,
\end{cases}
\end{equation}
where we note that
\begin{equation}
\kappa^2(k,Uk) = \frac{\epsilon \beta}{U}-\frac{\beta^2}{4}.
\end{equation}
Here $\kappa^2$ must be positive if any waves are generated by the vortex by \cref{eq:cond_wave}. As described in \cref{eq:dipole_bc}, far from the vortex, the vortex appears as a point dipole of strength $\mu$ hence we impose the boundary condition
\begin{equation}
\phi = \frac{\mu}{2} \delta(x) \quad \textrm{at} \quad y = 0^+.
\end{equation}
We also take $\phi \to 0$ as $y \to \infty$ and impose continuity of $\phi$ and $\phi_y+[H_y/(2H)]\phi$ across $y = D$ so that the velocity $(u,v)$ is continuous here. We note that this approach is equivalent to the matching step between an interior vortex and an exterior wave field in a full asymptotic expansion in small $\epsilon,\,\beta$  \citep{FLIERLHAINES94,JohnsonCrowe20,Croweetal20}.

We now express $\phi$ using a Fourier transform as
\begin{equation}
\label{eq:FT_def}
\phi(x,y) = \frac{1}{2\pi}\int_{-\infty}^\infty\widehat{\phi}(k,y)\exp(\rmi kx)\dint k,
\end{equation}
where $\widehat{\phi}$ satisfies the system
\begin{equation}
\left[\pder{^2}{y^2} - k^2\right] \widehat{\phi} = \begin{cases}
-\kappa^2\widehat{\phi} & y \leq D,\\
0 & y \geq D,
\end{cases}
\end{equation}
subject to
\begin{equation}
\begin{cases}
\left.\widehat{\phi}\right. = \frac{\mu}{2} & \textrm{at} \quad y = 0,\\
\left.\widehat{\phi}\right. \to 0           & \textrm{as} \quad y \to \infty,\\
\left[\widehat{\phi}\right]^+_- = 0  & \textrm{at} \quad y = D, \\
\left[\widehat{\phi}_y + \frac{H_y}{2H} \widehat{\phi} \right]^+_- = 0 & \textrm{at} \quad y = D,
\end{cases}
\end{equation}
where we note that $H_y/H = \beta$ for $y<D$ and $H_y/H = 0$ for $y > D$. The solution for $\widehat{\phi}$ is given by
\begin{equation}
\widehat{\phi} = \frac{\mu}{2}\begin{cases}
 \left(\widehat{C} \sin\left[ \sqrt{\kappa^2-k^2} \,y\right] + \cos\left[ \sqrt{\kappa^2-k^2} \,y\right]\right)& y \leq D,\\
\left(\widehat{C} \sin\left[ \sqrt{\kappa^2-k^2} D\right] + \cos\left[ \sqrt{\kappa^2-k^2} D\right]\right) e^{-|k|\left(y-D\right)} & y \geq D,
\end{cases}
\end{equation}
where
\begin{equation}
\widehat{C} = \frac{\sqrt{\kappa^2-k^2}\sin\left[ \sqrt{\kappa^2-k^2} D\right] - \left(|k|+\frac{\beta}{2}\right)\cos\left[ \sqrt{\kappa^2-k^2} D\right]}{\sqrt{\kappa^2-k^2}\cos\left[ \sqrt{\kappa^2-k^2} D\right]+\left(|k|+\frac{\beta}{2}\right)\sin\left[ \sqrt{\kappa^2-k^2} D\right]}.
\end{equation}
We note that the denominator of $\widehat{C}$ vanishes if the dispersion relation in \cref{eq:disp_rel} is satisfied therefore the wave modes correspond to the residues of these poles.

By the radiation condition that $c_g - U < 0$ we do not expect any waves generated upstream of the vortex and hence we only consider the solution far downstream where $x<0$ and $|x| \gg 1$. For large $x$, the exponential term in \cref{eq:FT_def} is strongly oscillatory and the terms without poles decay as $1/x$. We therefore consider only the terms containing $\widehat{C}$ and form a closed contour by including the arc $|k| = R$ with $R\to\infty$ in the lower half of the complex $k$ plane. All poles occur along the real line and are taken to lie within the contour. Finally, the contribution from the arc vanishes as $R\to \infty$ giving that
\begin{equation}
\phi \sim -\frac{\rmi\mu}{2}\sum \textrm{Res}[\widehat{\phi},k_n] \exp(\rmi k_n x),
\end{equation}
where $\textrm{Res}[f,x]$ denotes the residue of $f$ and $x$. Here we sum over all poles of $\widehat{\phi}$ and have gained an additional factor of $-1$ due to the orientation of the contour.

Since $\widehat{\phi}$ is even in $k$ there will be a pole at $-k = k_n$ for each pole at $k = k_n$ with residue of the opposite sign. We therefore have
\begin{equation}
\phi \sim \frac{\rmi\mu}{2}\sum_{n=1}^N \textrm{Res}[\widehat{\phi},k_n] \left[  \exp(-\rmi k_n x)-\exp(\rmi k_n x)\right] = \mu\sum_{n=1}^N \textrm{Res}[\widehat{\phi},k_n] \sin(k_n x),
\end{equation}
where the $k_n$ are the positive poles of $\widehat{C}$ and hence are the solutions to the dispersion relation corresponding to a mode of phase speed $U$. $N$ describes the number of modes for which $U = c_p$ and will be determined later. By differentiating the denominator of $\widehat{C}$ we find that the residues are given by
\begin{equation}
\textrm{Res}[\widehat{\phi},k_n] =  A_n \begin{cases} \sin\left[ \sqrt{\kappa^2-k_n^2} \,y\right] & y\leq D,\vspace{1pt}\\
\sin\left[ \sqrt{\kappa^2-k_n^2} D\right]e^{-k_n\left(y-D\right)} & y \geq D,
\end{cases}
\end{equation}
where
\begin{equation}
\label{eq:A_n_def}
A_n = \frac{l_n \left[ \epsilon + U k_n \right]}{k_n D \left[\epsilon+ U k_n\right] + \left[ \epsilon + \tfrac{U}{2}\left(k_n-\tfrac{\beta}{2}\right)\right]},
\end{equation}
for offshore wavenumber $l_n = \sqrt{\kappa^2-k_n^2}$ satisfying
\begin{equation}
\label{eq:l_n_eqn}
\tan(l_n D) = -\frac{l_n}{k_n+\tfrac{\beta}{2}}.
\end{equation}

The total wave field for large, negative $x$ is now given by
\begin{equation}
\label{eq:phi_sol}
\phi \sim \mu \begin{cases}
\sum_{n=1}^N  \left[A_n
\sin(l_n y) \sin(k_n x)\right] & y\leq D,\vspace{3pt}\\
\sum_{n=1}^N  \left[A_n \sin(l_n D)\sin(k_n x) \,e^{-k_n\left(y-D\right)}\right]  & y \geq D.
\end{cases}
\end{equation}
The values of $(k_n,l_n)$ can be determined numerically by finding the roots of $c_p = U$ using \cref{eq:omega_c_p} (with $l(k)$ given by \cref{eq:disp_rel}) or by solving \cref{eq:l_n_eqn} directly as a function of $k_n$. It may be shown that the modal components of $\phi$ are mutually orthogonal in the $y$ direction.

Finally, by calculating the maximum value of $c_p$ for each mode and comparing this to $U$, we may determine the total number modes, N, as the greatest integer such that
\begin{equation}
l_N \leq \kappa = \sqrt{\frac{\epsilon\beta}{U}-\frac{\beta^2}{4}}.
\end{equation}
Using the bounds on $l$ from \cref{eq:l_bound} we have
\begin{equation}
\label{eq:N_bound}
\left\lfloor \frac{D}{\pi}\sqrt{\frac{\epsilon\beta}{U}-\frac{\beta^2}{4}} \right\rfloor \leq N \leq \left\lfloor \frac{1}{2}+\frac{D}{\pi}\sqrt{\frac{\epsilon\beta}{U}-\frac{\beta^2}{4}} \right\rfloor,
\end{equation}
where $\lfloor * \rfloor$ denotes the `floor' function. By examining the form of \cref{eq:l_n_eqn}, we observe that if the value of $T = \tan(\kappa D) + 2\kappa/\beta$ is non-negative then equality holds in the upper bound of \cref{eq:N_bound} whereas if $T < 0$ then equality holds in the lower bound. The case of $N = 0$ corresponds to the vortex moving faster than the fastest wave and is equivalent to the second inequality in \cref{eq:cond_wave} not being satisfied.

\begin{figure}
	\centering
	\begin{subfigure}[b]{0.49\textwidth}
	\centering
	\includegraphics[trim={0 0 0 0},clip,width=\textwidth]{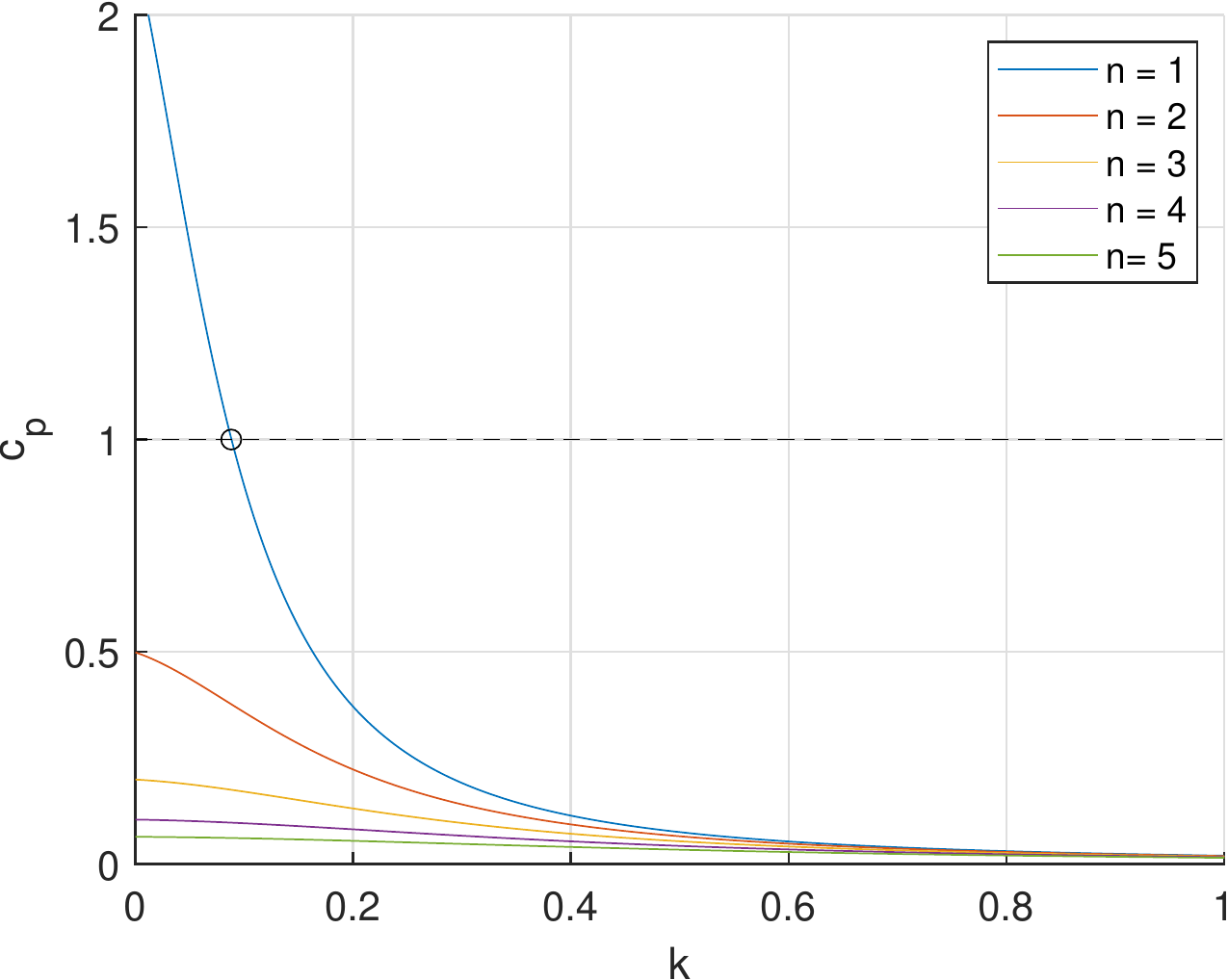}
	\caption{}
	\end{subfigure}
	\begin{subfigure}[b]{0.49\textwidth}
	\centering
	\includegraphics[trim={0 0 0 0},clip,width=\textwidth]{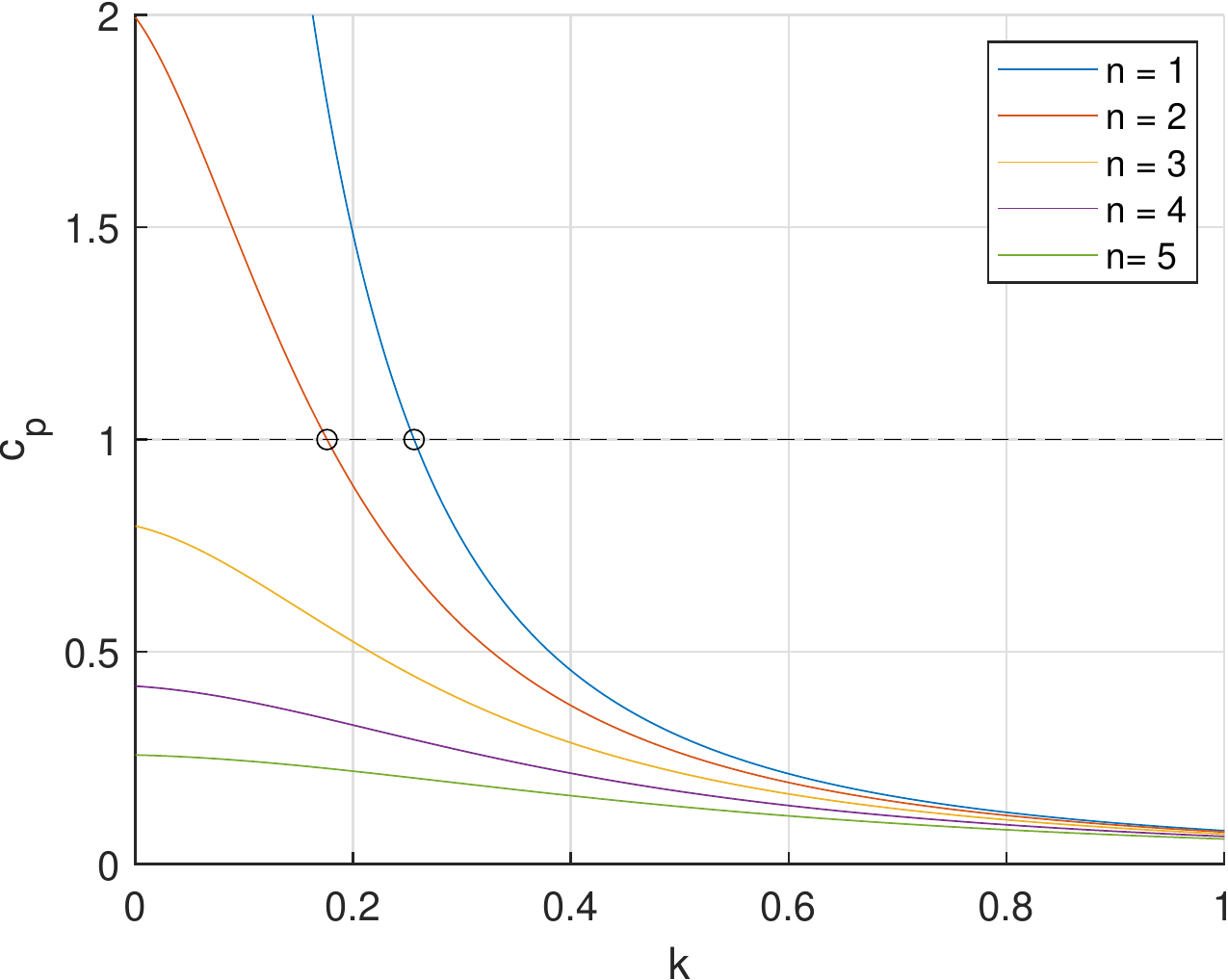}
	\caption{}
	\end{subfigure}
	\caption{Plots of the phase speed, $c_p$, for the first five modes with $\beta = 0.1$, $D = 25.6$.  (a)  $\epsilon = 0.2$.  (b)  $\epsilon = 0.8$. The dotted line denotes $U = 1$ and the $k_n$ are determined as the intersections $c_p = U$ and denoted by the open circles. We observe one mode for $\epsilon = 0.2$ and two modes for $\epsilon = 0.8$ for this choice of parameters.}
    \label{fig:fig3}
\end{figure}

\cref{fig:fig3} shows the solutions of $c_p = U$ for $\beta = 0.1$, $D = 25.6$, $U = 1$ and $\epsilon \in\{ 0.2, 0.8\}$. The alongshore wavenumber, $k_n$, for which a given mode has $c_p = U$ is shown by an open circle. The value of $l_n$ can be easily determined using $l_n = \sqrt{\kappa^2-k_n^2}$. We note that if a vortex slows down, it will generate an increased number of modes as $\kappa$ will increase.

\subsection{Wave energy flux and vortex decay}

As the vortex generates waves, it loses energy to the wave-field and decays. Since the group velocity for all waves is negative in the frame of the vortex, all energy emitted will cross the line $x = -L$ for $L \gg 1$ where the wave-field is small amplitude and hence linear to leading order. Therefore, this energy flux is given to leading order by the quadratic pressure work plus the transport of energy across the line due to the moving coordinates \citep{Croweetal20}; so
\begin{equation}
F = \int_0^\infty H\left[  -pu + \tfrac{1}{2}\left(u^2+v^2\right)U\right] \dint y,
\end{equation}
where the factor of $H$ is obtained by integrating over the layer depth. For linear waves, the pressure may be determined from \cref{eq:gov_eq} as
\begin{equation}
p = Uu+\frac{\epsilon}{H}\psi,
\end{equation}
and hence
\begin{equation}
\label{eq:F_psi}
F =\int_0^\infty \left\{\frac{U}{2H}\left[\left(\pder{\psi}{x}\right)^2-\left(\pder{\psi}{y}\right)^2\right]+\frac{\epsilon\psi}{H}\pder{\psi}{y} \right\} \dint y.
\end{equation}

\cref{eq:F_psi} may be written in terms of $\phi = \psi/\sqrt{H}$ as
\begin{equation}
\label{eq:F_phi}
F =-\int_0^\infty \left[ \left(\frac{U H_y^2}{8H^2}-\frac{\epsilon H_y}{2 H}\right) \phi^2 + \left(\frac{U H_y}{2H} - \epsilon \right) \phi \phi_y + \frac{U}{2}\phi_y^2 - \frac{U}{2} \phi_x^2\right] \dint y,
\end{equation}
and calculating $F$ using \cref{eq:phi_sol} gives
\begin{equation}
F = \frac{U\mu^2}{4}\sum_{n=1}^N \frac{k_n A_n^2}{\epsilon+U k_n} \left( \left[ \epsilon + \tfrac{U}{2}\left(k_n-\tfrac{\beta}{2}\right)\right] + k_n D \left[\epsilon+ U k_n\right]\right).
\end{equation}
which we note is independent of $x$. Substituting for $A_n$ using \cref{eq:A_n_def} we have
\begin{equation}
\label{eq:F_sol}
F = \frac{U\mu^2}{4}\sum_{n=1}^N \frac{k_n l_n^2 \left[\epsilon+ U k_n\right]}{\left[ \epsilon + \tfrac{U}{2}\left(k_n-\tfrac{\beta}{2}\right)\right] + k_n D \left[\epsilon+ U k_n\right]},
\end{equation}
which we note is always positive corresponding to a loss of vortex energy. Equating this flux with the loss of vortex energy, $E$, gives
\begin{equation}
\label{eq:E_evol}
\frac{dE}{dt} = -F.
\end{equation}
If we now assume that the vortex remains self similar throughout the evolution, the energy, $E$, and dipole strength, $\mu$, may be determined in terms of the vortex speed $U$ and radius $a$. We now have two quantities, $U$ and $a$, with a single evolution equation, \cref{eq:E_evol}, so a second equation is required to close the system. Following \citet{FLIERLHAINES94}, \citet{JohnsonCrowe20} and \citet{Croweetal20} we choose conservation of centre vorticity so that the maximum vorticity of the vortex remains constant throughout the evolution. This gives a second equation
\begin{equation}
\label{eq:centre_vor}
\frac{d\eta_c}{dt} = 0,
\end{equation}
where $\eta_c = \eta_c(U,a)$ is the maximum vorticity within the vortex can be determined from the vortex solution, either numerically or analytically.

In the case of asymptotically small $\beta$ and $\epsilon$ the vortex solution reduces the the classical Lamb-Chaplygin dipole and the quantities may be determined analytically from \cref{eq:LCD} as
\begin{equation}
\label{eq:asymp_quant}
E(U,a) = \pi U^2 a^2,\quad \mu(U,a) = 2\pi U a^2, \quad \textrm{and} \quad \eta_c(U,a) \propto \frac{U}{a}.
\end{equation}
Therefore \cref{eq:E_evol,eq:centre_vor} give
\begin{equation}
\label{eq:energy_asymp}
\frac{d}{dt}(U^2 a^2) = -\pi U^3 a^4 \sum_{n = 1}^N \frac{k_n l_n^2 \left[\epsilon+ U k_n\right]}{\left[ \epsilon + \tfrac{U}{2}\left(k_n-\tfrac{\beta}{2}\right)\right] + k_n D \left[\epsilon+ U k_n\right]},
\end{equation}
and
\begin{equation}
\label{eq:peak_vor}
\frac{d}{dt}\left(\frac{U}{a}\right) = 0.
\end{equation}
Therefore $a(t) \propto U(t)$ so \cref{eq:energy_asymp} may be solved as an equation for $U(t)$ subject to some initial condition
\begin{equation}
(U,a) = (U_0,a_0)\quad \textrm{at}\quad t = t_0.
\end{equation}
We note that since the wavevector, $(k_n,l_n)$, and number of modes, $N$, both have a complicated dependence on $U$ this equation would have to be solved numerically.

\begin{figure}
	\centering
	\begin{subfigure}[b]{0.50\textwidth}
	\centering
	\includegraphics[trim={0 0 0 0},clip,width=\textwidth]{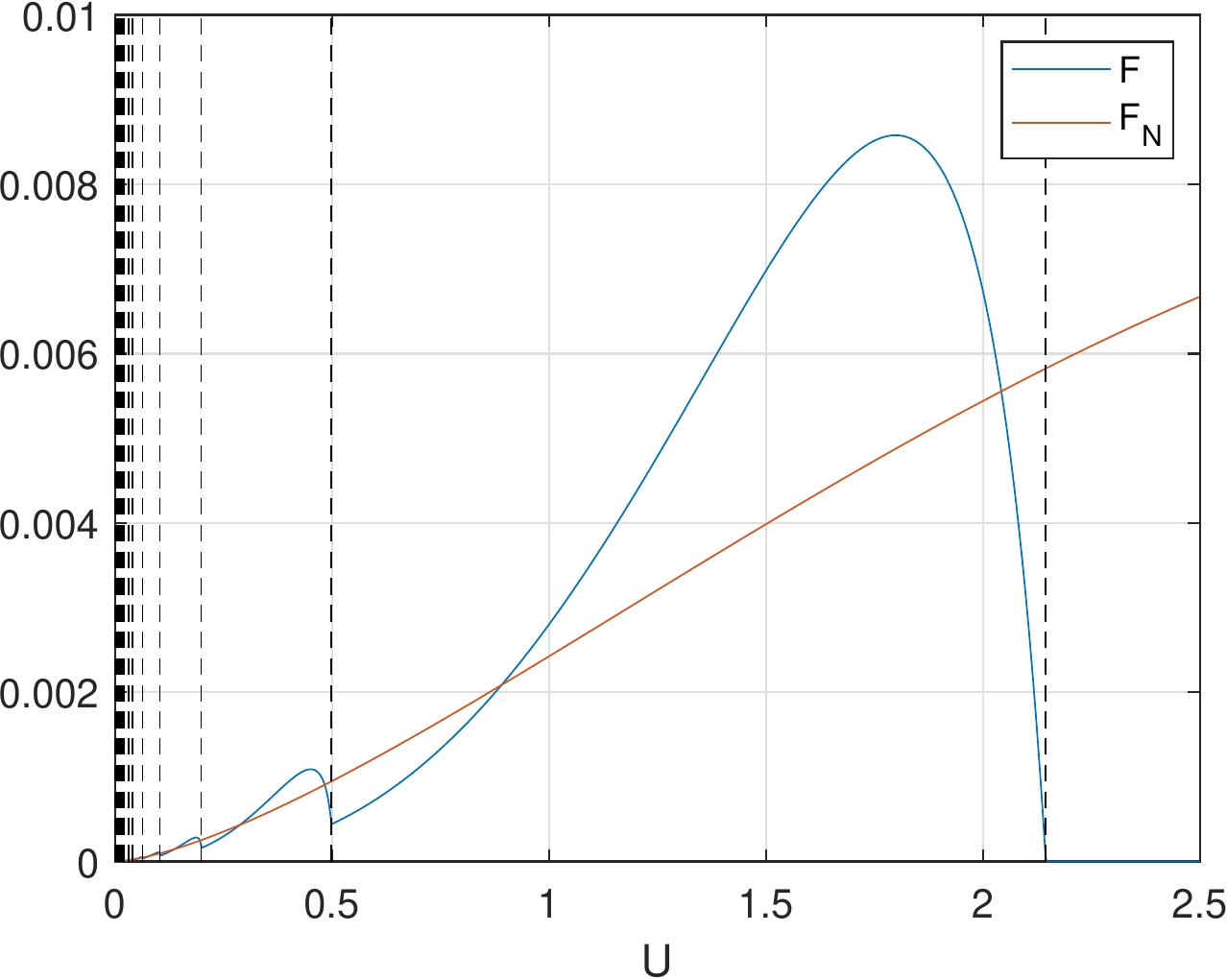}
	\caption{}
	\end{subfigure}
	\begin{subfigure}[b]{0.48\textwidth}
	\centering
	\includegraphics[trim={0 0 0 0},clip,width=\textwidth]{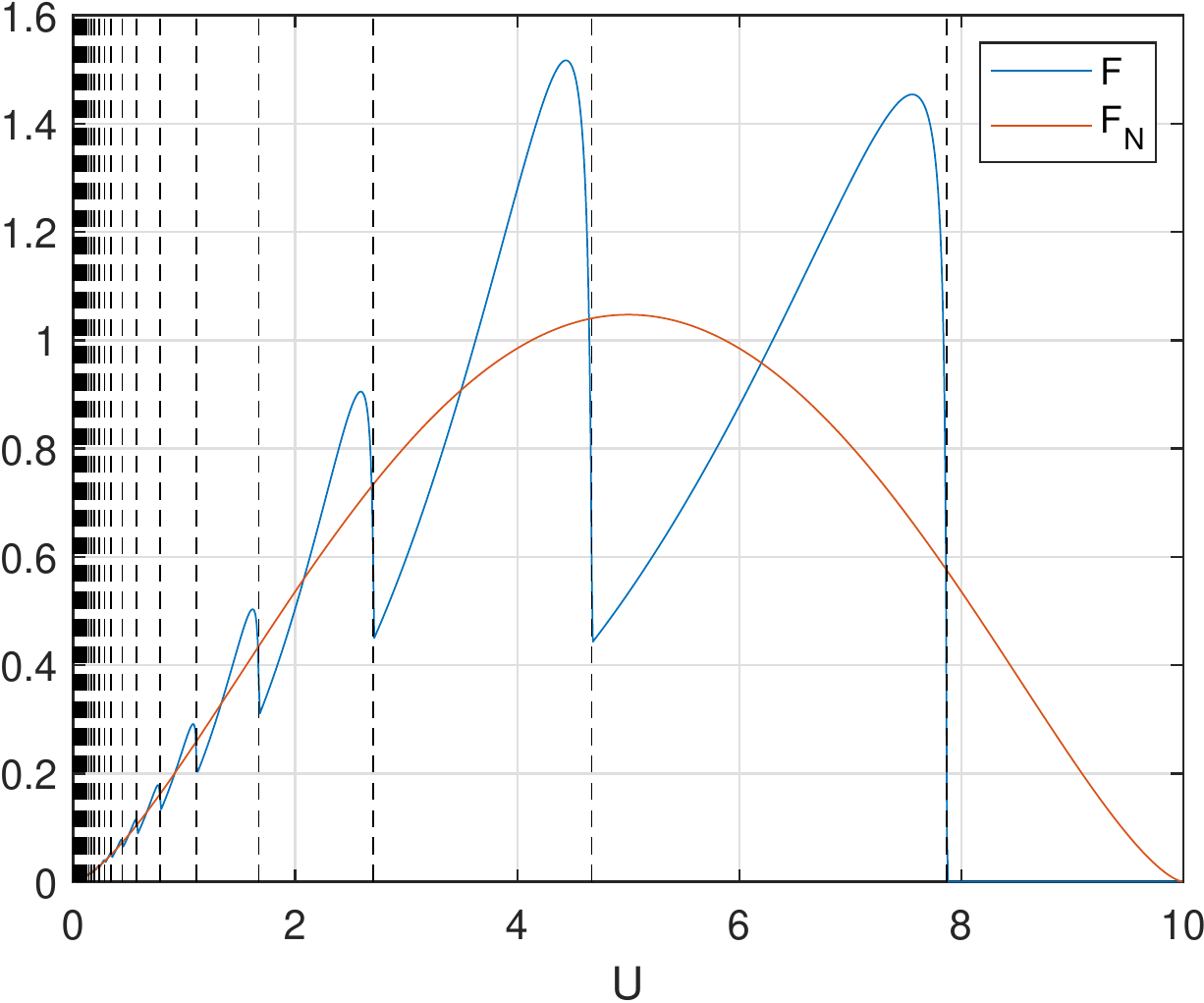}
	\caption{}
	\end{subfigure}
	\caption{Plots of the wave energy flux $F$ (blue) and the limit of $F$ as $N \to \infty$, denoted $F_N$ (red). $F$ and $F_N$ are shown as a function of $U$ for $(\epsilon,\beta) = (0.2,0.1)$ (a) and $(1,0.4)$ (b) with $D = 25.6$ and $\mu = 2\pi U$. If the vortex speed, $U$, exceeds the fastest wave there is no wave energy flux, this occurs for $U > 2.14$ for panel (a) and for $U>7.87$ for panel (b).}
    \label{fig:fig4}
\end{figure}

\cref{fig:fig4} shows the wave energy flux, $F$, as a function of $U$ for $(\epsilon,\beta) = (0.2,0.1)$ and $(1,0.4)$ with $D = 25.6$ and $\mu = 2\pi U$. For large values of $U$ the wave energy flux vanishes as there are no modes which match the vortex speed. As we decrease $U$, an increasing number of modes can satisfy $c_p = U$ so new modes appear in our solution. The dashed lines in \cref{fig:fig4} denote the values of $U$ at which these modes appear (disappear) as $U$ is decreased (increased). Whenever a new mode appears, a peak corresponding to this mode is seen in the energy flux similar to the results of \citet{JOHNSON79}. New modes appear with $k_n = 0$ (where $c_p$ is maximal) then $k_n$ increases as $U$ decreases with the associated energy flux moving through a maximum and dropping off. We now consider the limit of small $U$ where the number of modes, $N$ becomes large.

\subsubsection{The large $N$ limit}

In the case of a large number of modes, we must have that
\begin{equation}
N \approx \frac{D}{\pi} \sqrt{\frac{\epsilon\beta}{U}-\frac{\beta^2}{4}} \gg 1.
\end{equation}
Therefore this limit occurs if the velocity is small ($U \ll 1$) or the shelf width is large ($D \gg 1$). Noting that $N \approx D\kappa/\pi$ and $k_n = O(\kappa)$ we have that $k_n D$ is large and hence, from \cref{eq:F_sol}, we have
\begin{equation}
F \approx \frac{U\mu^2}{4} \sum_{n=1}^N \frac{l_n^2}{D}.
\end{equation}
For large $n$, the offshore wavenumber $l_n$ may be approximated using \cref{eq:l_bound} as
\begin{equation}
l_n \approx \frac{n\pi}{D},
\end{equation}
so
\begin{equation}
F \approx \frac{U\mu^2}{4} \sum_{n=1}^N \frac{n^2\pi^2}{D^3} \approx \frac{U\mu^2}{12\pi} \left[\frac{\epsilon\beta}{U}-\frac{\beta^2}{4}\right]^{\frac{3}{2}},
\end{equation}
where we have used
\begin{equation}
\sum_{n=1}^N n^2 \approx \frac{1}{3} N^3,
\end{equation}
for large $N$. We now define $F_N$ to be the asymptotic form of $F$ for large $N$ so
\begin{equation}
\label{eq:F_N}
F_N = \frac{U\mu^2}{12\pi} \left[\frac{\epsilon\beta}{U}-\frac{\beta^2}{4}\right]^{\frac{3}{2}},
\end{equation}
where $F_N$ is plotted in \cref{fig:fig4} and can be seen to well describe $F$ for small $U$. Additionally, we observe that $F_N$ provides a fairly good approximation to $F$ for order one values of $U$.

Taking $\epsilon$ and $\beta$ to be small we may use \cref{eq:asymp_quant} to obtain the approximate evolution equations
\begin{equation}
\label{eq:E_large_N}
\frac{d}{dt}\left(U^2 a^2\right) = -\frac{U^3a^4}{3} \left[\frac{\epsilon\beta}{U}-\frac{\beta^2}{4}\right]^{\frac{3}{2}},
\end{equation}
and
\begin{equation}
\label{eq:q_large_N}
\frac{d}{dt}\left(\frac{U}{a}\right) = 0,
\end{equation}
which may be easily solved in the case of $4\epsilon/U \gg \beta$ for vortex speed and radius
\begin{equation}
\label{eq:U_a_evol_limit}
\left( U, a\right) = \left(U_0, a_0 \right) \left[ 1 + \frac{1}{8}\sqrt{\epsilon^3\beta^3 a_0^4/U_0}\, (t-t_0) \right]^{-\frac{2}{3}}.
\end{equation}
This solution describes a polynomial decay of the vortex speed and radius similar to the case of a beta plane modon considered by \citet{FLIERLHAINES94} and \citet{JohnsonCrowe20}. Further, we note that \cref{eq:E_large_N,eq:q_large_N} exactly correspond to the vortex decay in the continuous limit of an unbounded shelf, $D = \infty$, using the method of \citet{JohnsonCrowe20} and \citet{Croweetal20}. While the solution in \cref{eq:U_a_evol_limit} does require $4\epsilon/U \gg \beta$, it can be seen that if this condition is initially satisfied then it will remain true as $U$ decreases.

\section{Numerical simulations}
\label{sec:num_sim}

To test our predictions we perform numerical simulations using Dedalus \citep[setup file available as supplementary material]{BurnsVOLB20}. We solve the full nonlinear, rotating shallow water equations under the rigid lid assumption (see \cref{eq:gov_eq}) in a frame moving with constant speed, $U_f$, in the along-shore ($x$) direction. We use the numerical domain $(x, y) \in [-51.2, 51.2]\times [0,51.2]$ with $1024$ gridpoints in each direction and decompose fields in terms of a Fourier basis in the $x$ direction and a compound Chebyshev basis in the $y$ direction with separate Chebyshev expansions on and off the shelf. Solutions are integrated for $t \in [0,50]$ using a second order semi-implicit BDF scheme with a timestep of $10^{-3}$. We take boundary conditions of no flow through the walls at $y = 0$ and $y = 51.2$ and include small viscous terms with a viscosity of $\nu = 1.8\times 10^{-5}$ for numerical stability. The inclusion of viscosity requires additional boundary conditions so we impose free slip conditions on the walls, $\partial_y u = 0$ on $y = 0,\, 51.2$, and note that the leading order vortex solution, \cref{eq:LCD}, also satisfies these conditions. Therefore there is unlikely to be significant vorticity generation at the boundaries, something which can lead to a modification or breakdown of the vortex and is particularly prevalent using no-slip boundary conditions.

The use of a Fourier basis in the $x$ direction results in a periodic boundary and hence waves may loop around the domain and interfere with the vortex. However, the stop time, $t = 50$, is found to be sufficiently early that these waves do not interact with the vortex. Similarly, the solid wall at $y = 51.2$ differs from the semi-infinite domain used in our theoretical calculations. Since wavelike disturbances will decay exponentially off the shelf (in the region $y > D$), we'd expect any effects of this rigid wall to be exponentially small.

For all simulations the shelf slope, $\beta$, and shelf width, $D$, are chosen as $(\beta,D) = (0.1, 25.6)$. Simulations are initialised using the velocity fields corresponding to a Lamb-Chaplygin dipolar vortex (see \cref{eq:LCD}) with initial speed $|U(0)| = 1$ and radius $a(0) = 1$. The frame speed, $U_f$, is set to match the speed of this initial vortex. Therefore, we expect the vortex to remain close to $x = 0$ throughout the evolution with deviations occurring as the vortex speed changes.

The effects of non-zero $\beta$ and $\epsilon$ are to modify the initial vortex leading to a transient adjustment phase at the beginning of the simulation where the vortex adjusts to the effects of rotation and shelf slope and the wave field begins to develop. For small $\epsilon$ and $\beta$, this adjustment is small and the vortex remains approximately a Lamb-Chaplygin dipole with a modified speed and radius. In order to compare with our theoretical predictions, we take the values of $U_0$ and $a_0$ to be the speed and radius after this adjustment phase with $t = t_0$ describing the time taken for this adjustment to occur. A value of $t_0 = 2$ is found to be sufficient and comparison is made with the theory for $t \geq t_0$. We note that accurately determining the values of $U_0$ and $a_0$ from the numerical data is difficult. This can present issues when comparing with our theoretical predictions due to the sensitive dependence of \cref{eq:energy_asymp} on these quantities. The vortex speed, $U_0$, is determined by tracking the position of the vorticity maximum and the vortex radius, $a_0$, is estimated using the point at which the vorticity becomes $2\%$ of its maximum value.

\begin{figure}
	\centering
	\begin{subfigure}[b]{0.49\textwidth}
	\centering
	\includegraphics[trim={0 0 0 0},clip,width=\textwidth]{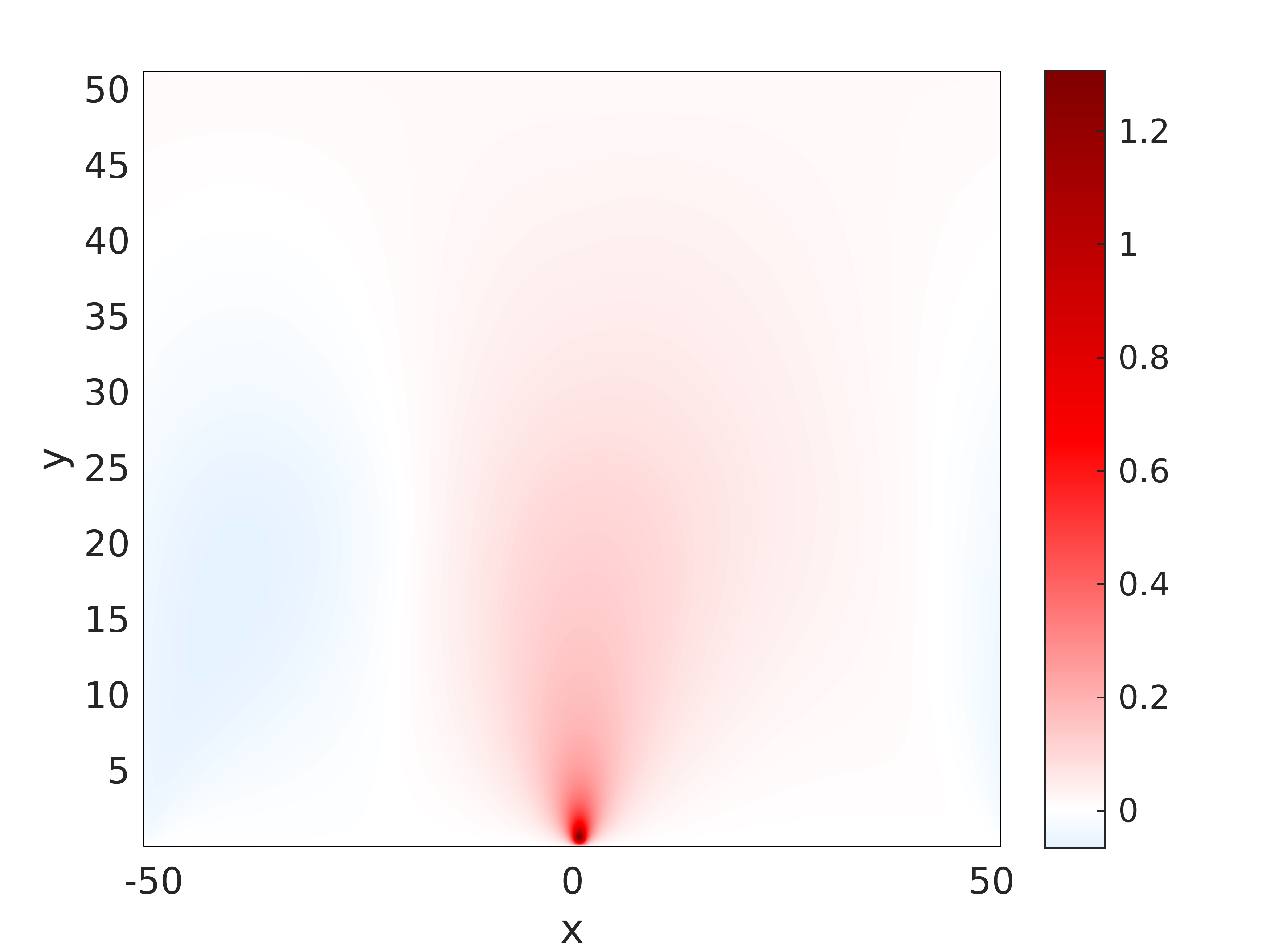}
	\caption{}
	\end{subfigure}
	\begin{subfigure}[b]{0.49\textwidth}
	\centering
	\includegraphics[trim={0 0 0 0},clip,width=\textwidth]{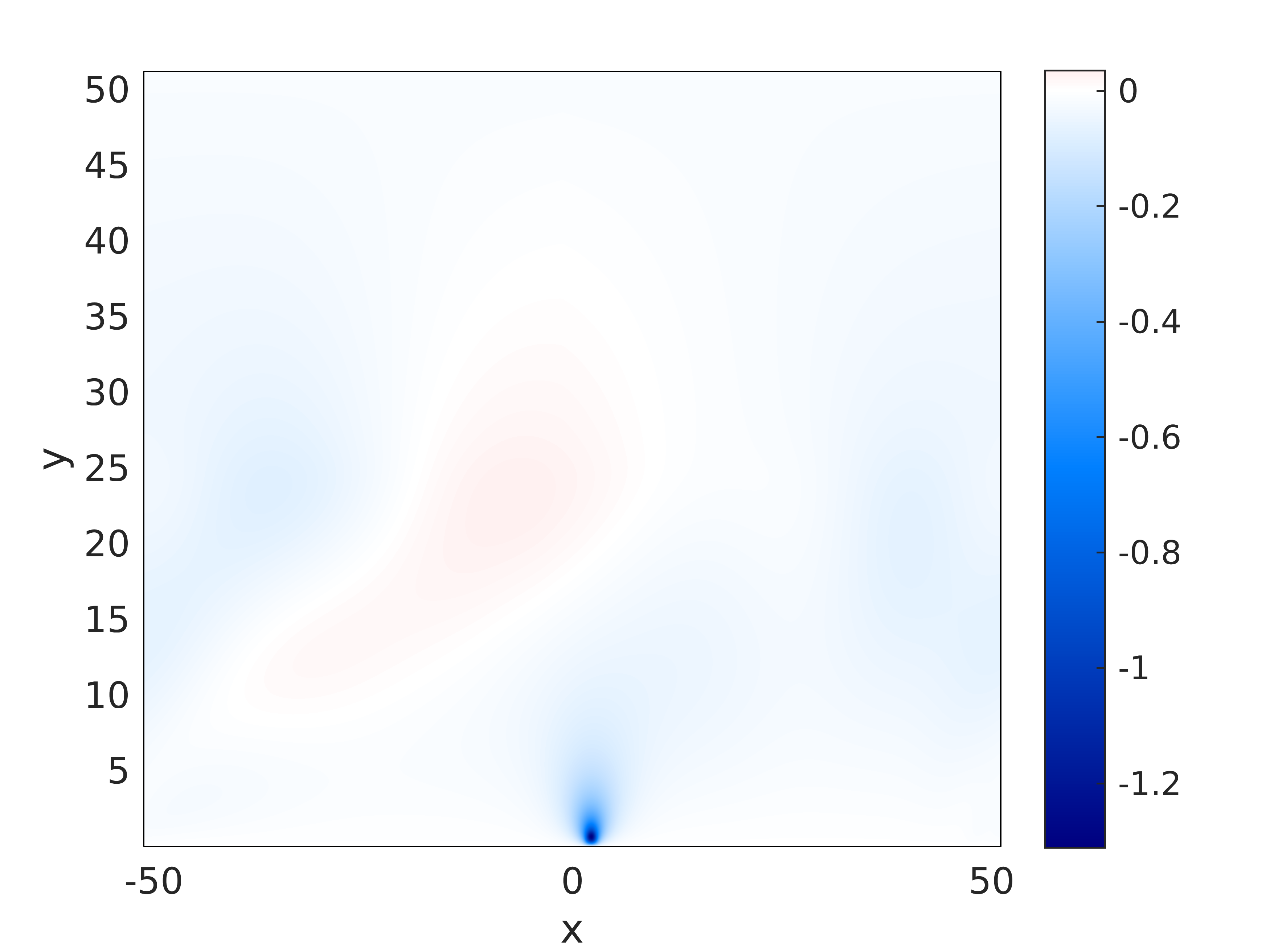}
	\caption{}
	\end{subfigure}
	\newline
	\begin{subfigure}[b]{0.49\textwidth}
	\centering
	\includegraphics[trim={0 0 0 0},clip,width=\textwidth]{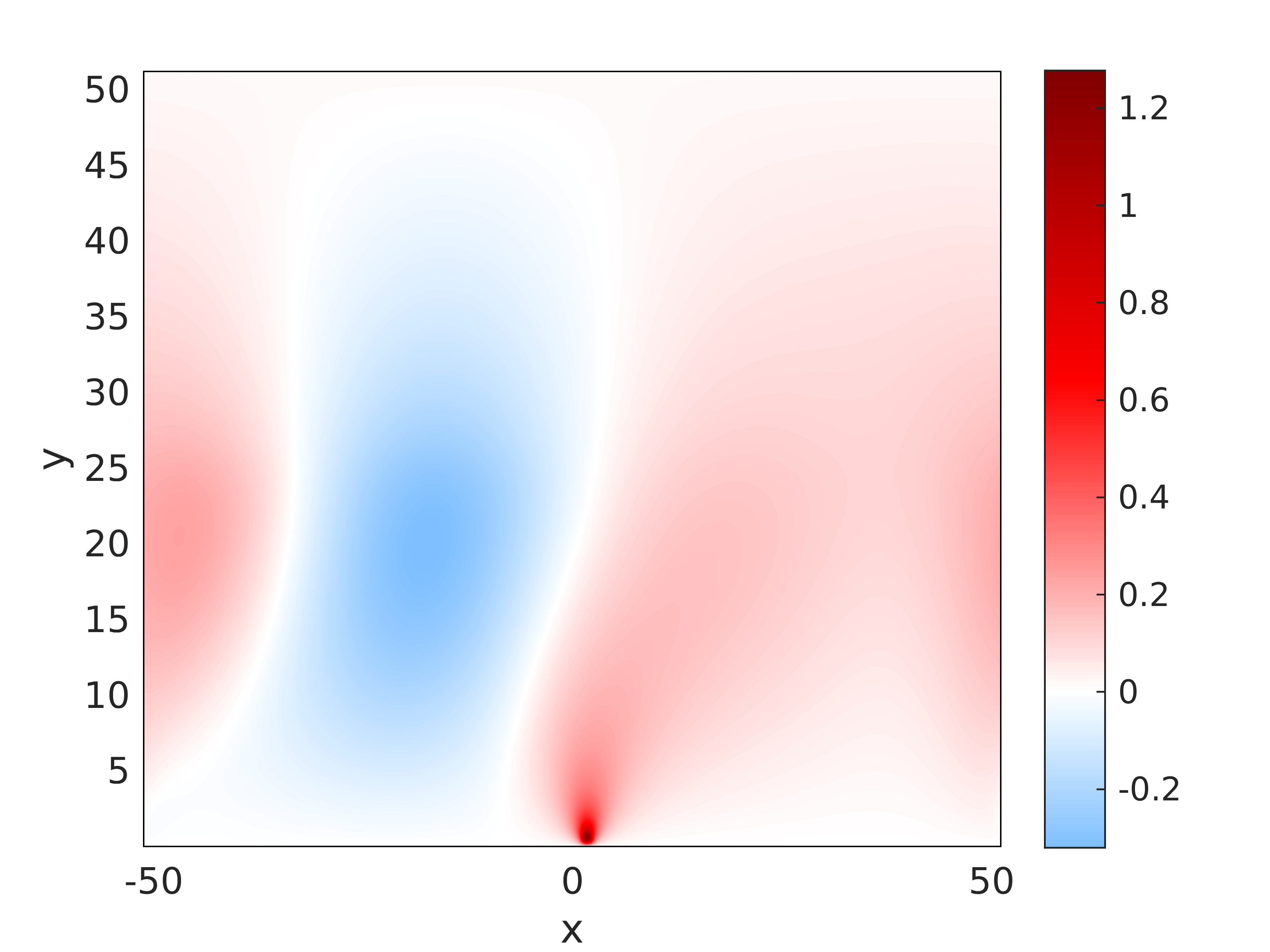}
	\caption{}
	\end{subfigure}
	\begin{subfigure}[b]{0.49\textwidth}
	\centering
	\includegraphics[trim={0 0 0 0},clip,width=\textwidth]{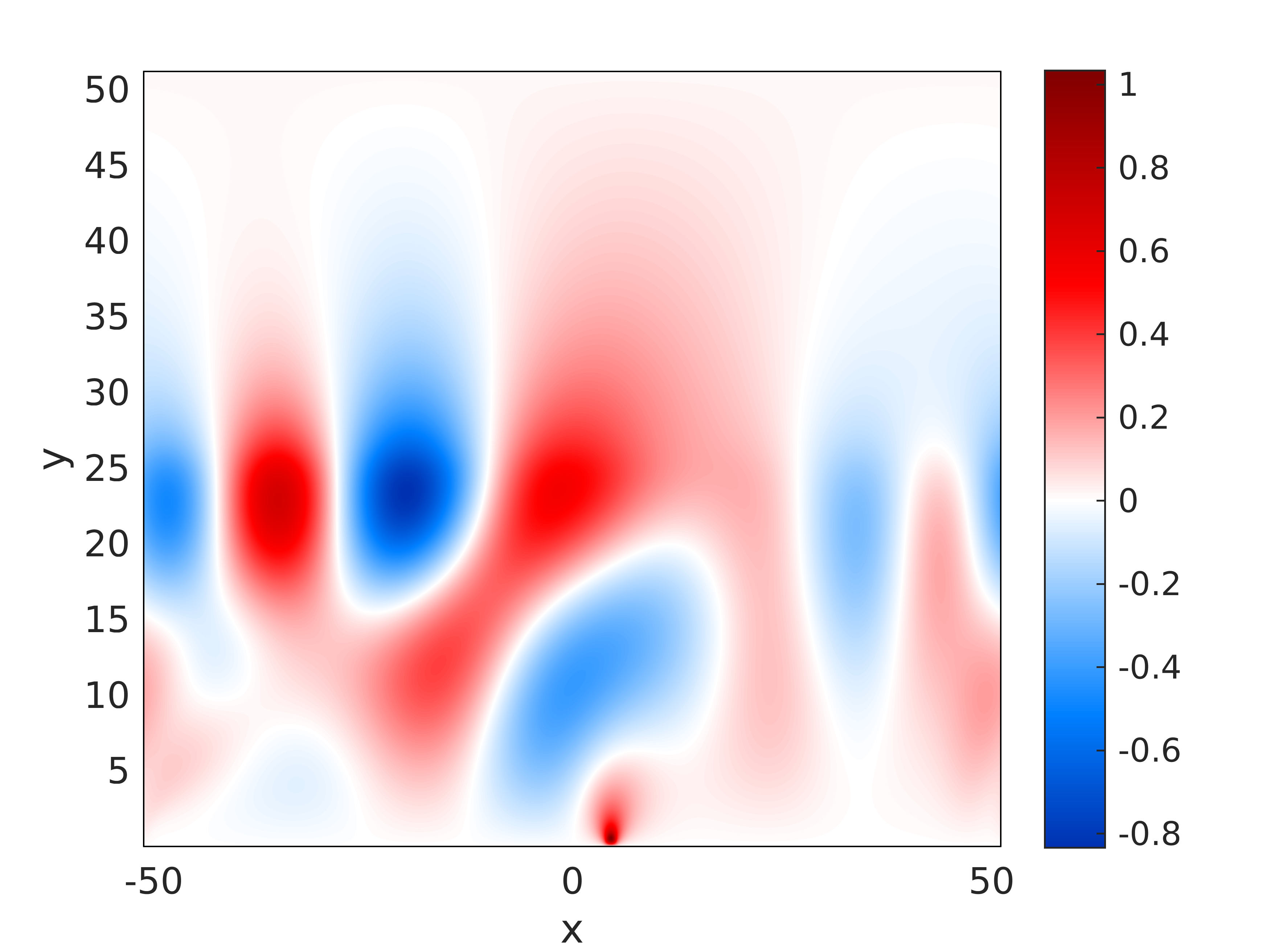}
	\caption{}
	\end{subfigure}
	\caption{The streamfunction, $\psi$, as a function of position in a frame moving with the speed of the initial vortex, $U_f = U(0)$. Results are shown for $\beta = 0.1$, $D = 25.6$ and $t = 50$, for various inverse Rossby numbers, $\epsilon$, and initial vortex speeds, $U(0)$. The vortex adjusts slightly due to finite $\epsilon$ and $\beta$ effects so the value of $U = U_0$ taken at $t = t_0 = 1$ can differ from $U(0)$ by up to $10-15\%$. (a) $(\epsilon,U(0)) = (0.05,1)$, a vortex travelling faster than all shelf waves. (b) $(\epsilon,U(0)) = (0.6,-1)$, a vortex moving in the opposite direction to all shelf waves. (c) $(\epsilon,U(0)) = (0.2,1)$, here vortex speed matches a single wave. (d) $(\epsilon,U(0)) = (1,1)$, here the vortex speed matches two waves.}
    \label{fig:fig5}
\end{figure}

\cref{fig:fig5} shows the streamfunction, $\psi$, for the final timestep, $t = 50$, of our numerical simulations for a range of parameters. Panels (a) and (b) show vortices which are respectively moving faster than and in the opposite direction to all shelf wave modes. For these simulations, the value of $\psi_c$ is conserved to within the error expected due to viscous effects and while a very weak wave signature is observed, this is likely the result of transient waves generated during the initial adjustment. \cref{fig:fig5}.(c) shows $\psi(x,y,50)$ for $U(0) = 1$ and $\epsilon = 0.2$ which initially matches the speed of a single wave with predicted wavelength of $\lambda_n = 2\pi/k_n = 71.2$. This wavenumber approximately matches the observed wave which we note is likely to be restricted by the length of the domain. Finally \cref{fig:fig5}.(d) shows $\psi(x,y,50)$ for $U(0) = 1$ and $\epsilon = 1$. The initial speed, $U(0) = 1$, is very close to matching the phase speed of the first three modes, however the vortex undergoes significant adjustment due to the fairly large value of $\epsilon$ and adjusts to a value of $U_0 \approx 1.1$. This value of $U_0$ only matches the speed of the first two modes and, according to our theoretical predictions, corresponds to modes with wavelengths of $\lambda_n = 22.7$ and $\lambda_n = 31.0$ and offshore wavenumbers of $l_n = 0.11$ and $l_n = 0.22$ respectively. This prediction appears consistent with our simulation where both these modes are seen. We note that as the vortex speeds change throughout the evolution, so to will the wavenumbers of the generated mode. The number of generated modes may also change if the vortex speed slows sufficiently to excite a new mode.

For small $\epsilon$ and $\beta$ we can show using the Lamb-Chaplygin solution that the total streamfunction at the position of the maximum vorticity is proportional to $Ua$. Therefore, assuming that \cref{eq:peak_vor} holds, we have
\begin{equation}
\label{eq:psi_c}
\frac{\psi_c(t)}{\psi_0} = \frac{U(t)\, a(t)}{U_0 a_0} = \frac{U^2(t)}{U_0^2},
\end{equation}
where $\psi_c$ is the value of the streamfunction at the position of maximum vorticity and $\psi_0$ is the value of $\psi_c$ at $t = t_0 = 2$. The normalised value of $\psi_c$ can be easily determined from our simulations and used to test our decay predictions by comparing with solutions of \cref{eq:energy_asymp,eq:peak_vor} as well as the polynomial decay prediction in \cref{eq:U_a_evol_limit}.

For very small values of $\epsilon$ and $\beta$ the wave energy flux, $F$, is small such that the vortex decay, and hence the decrease in $\psi_c$, is slow. Since the effect of viscosity is to decrease the domain averaged energy by around $1-2\%$ over the time interval $t \in [0,50]$, it is not possible to accurately determine how the wave energy flux affects the evolution of the vortex energy when the energy lost to wave field is similar to the viscous dissipation. Conversely, for values of $\epsilon$ greater that $1$, while the wavefield remains small due to small $\beta$, the vortex is no longer well described by the Lamb-Chaplygin solution and the asymptotic expressions for $E$, $\mu$ and $\eta_c$ in \cref{eq:asymp_quant} begin to deviate from the true values. Though these deviations are fairly small despite an order $1$ value of $\epsilon$, we find that \cref{eq:E_evol} is very sensitive to the values of $E$ and $\mu$ and our prediction gives only the order of magnitude of the decay scale rather than an accurate result. As a compromise between these limits we consider here the cases of $\epsilon \in \{0.4, 0.6, 0.8\}$, $\beta = 0.1$ and $(U(0),a(0)) = (1,1)$ which we observe are well described by the Lamb-Chaplygin solution. As describes above, the values of $U_0$ and $a_0$ are predicted from the vortex speed and radius at $t = t_0$.

\begin{figure}
	\centering
	\begin{subfigure}[b]{0.49\textwidth}
	\centering
	\includegraphics[trim={0 0 0 0},clip,width=\textwidth]{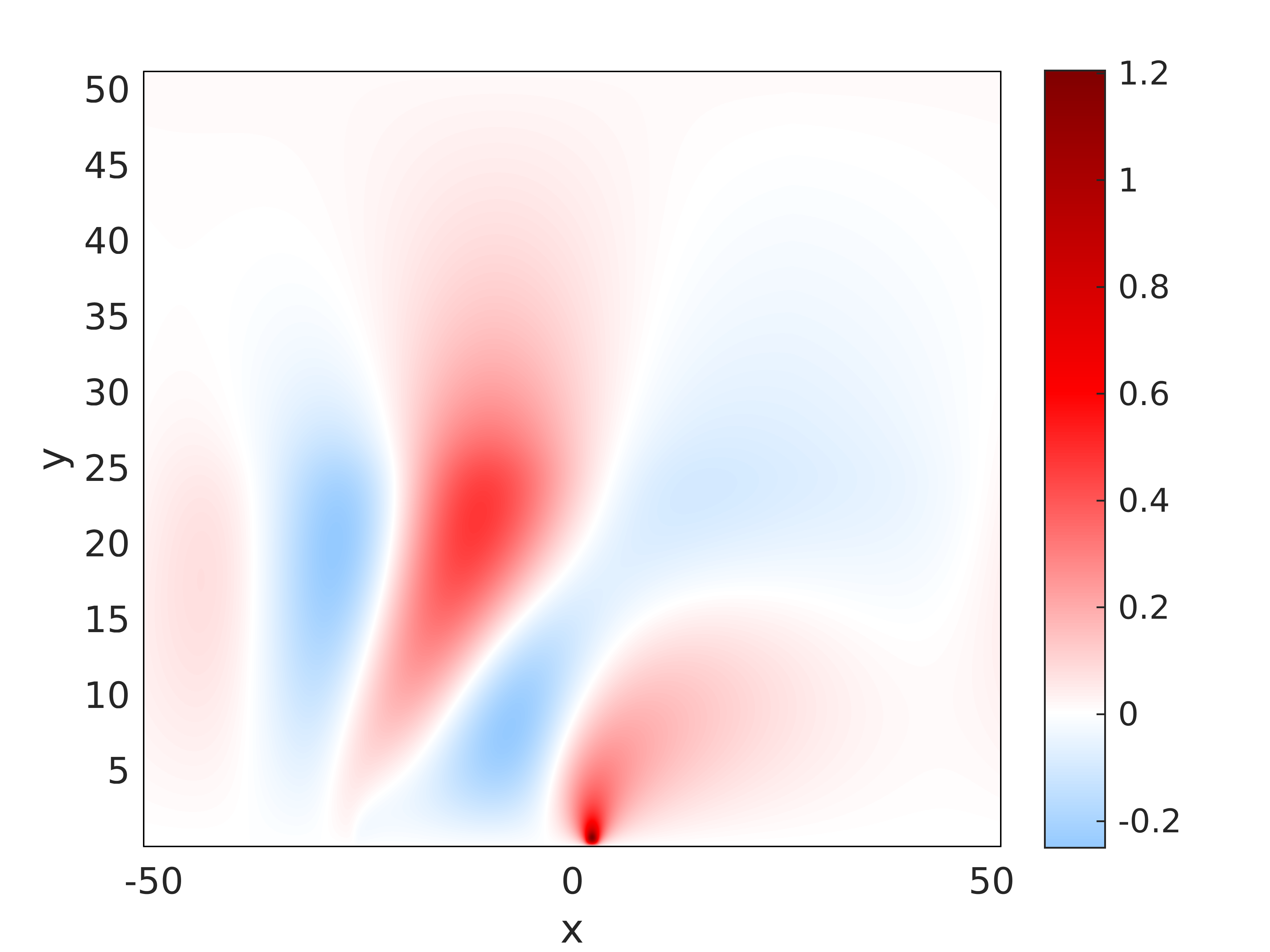}
	\caption{}
	\end{subfigure}
	\begin{subfigure}[b]{0.49\textwidth}
	\centering
	\includegraphics[trim={0 0 0 0},clip,width=\textwidth]{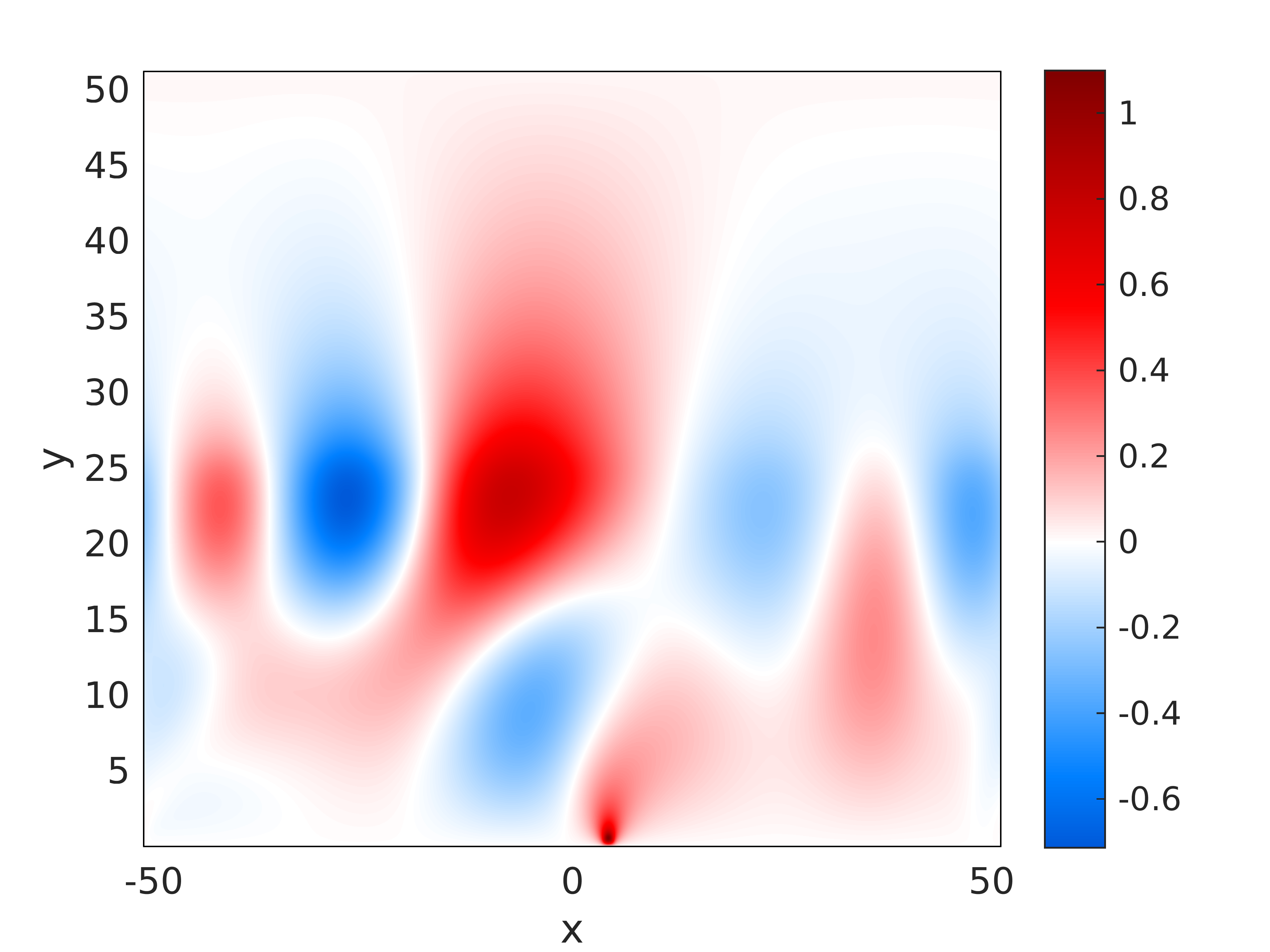}
	\caption{}
	\end{subfigure}
	\caption{Plots of the streamfunction, $\psi$, showing the formation of the wavefield for $\epsilon = 0.8$, $U_0 = 1.15$, $a_0 = 1$, $\beta = 0.1$ and $D = 25.6$ at two values of $t$, $t = 25$ (a) and $t = 50$ (b). Similarly to \cref{fig:fig5}, the solution is shown in a frame moving with speed $U_f = 1$ in the $x$ direction.}
    \label{fig:fig6}
\end{figure}

\begin{figure}
	\centering
	\begin{subfigure}[b]{0.49\textwidth}
	\centering
	\includegraphics[trim={0 0 0 0},clip,width=\textwidth]{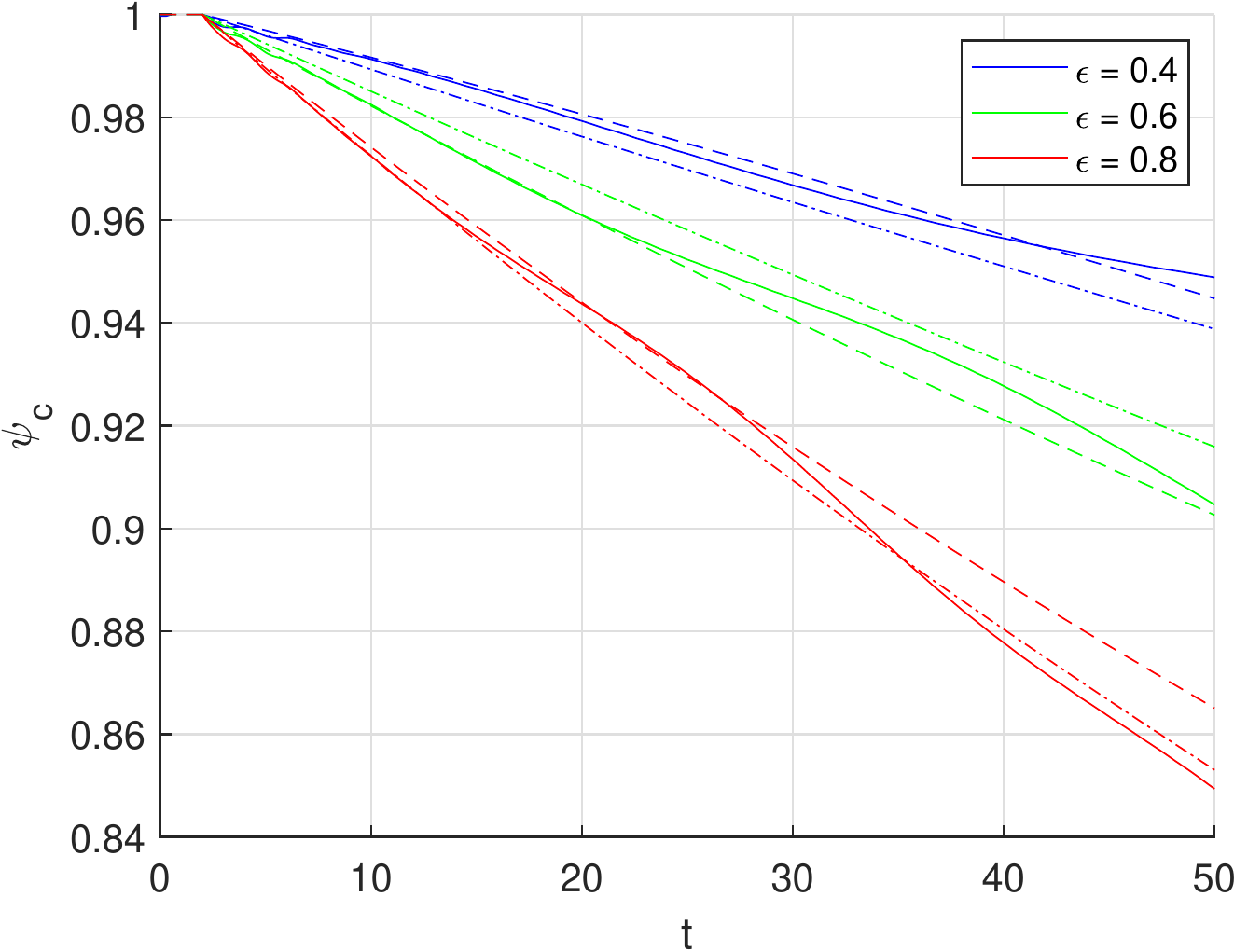}
	\caption{}
	\end{subfigure}
	\begin{subfigure}[b]{0.49\textwidth}
	\centering
	\includegraphics[trim={0 0 0 0},clip,width=\textwidth]{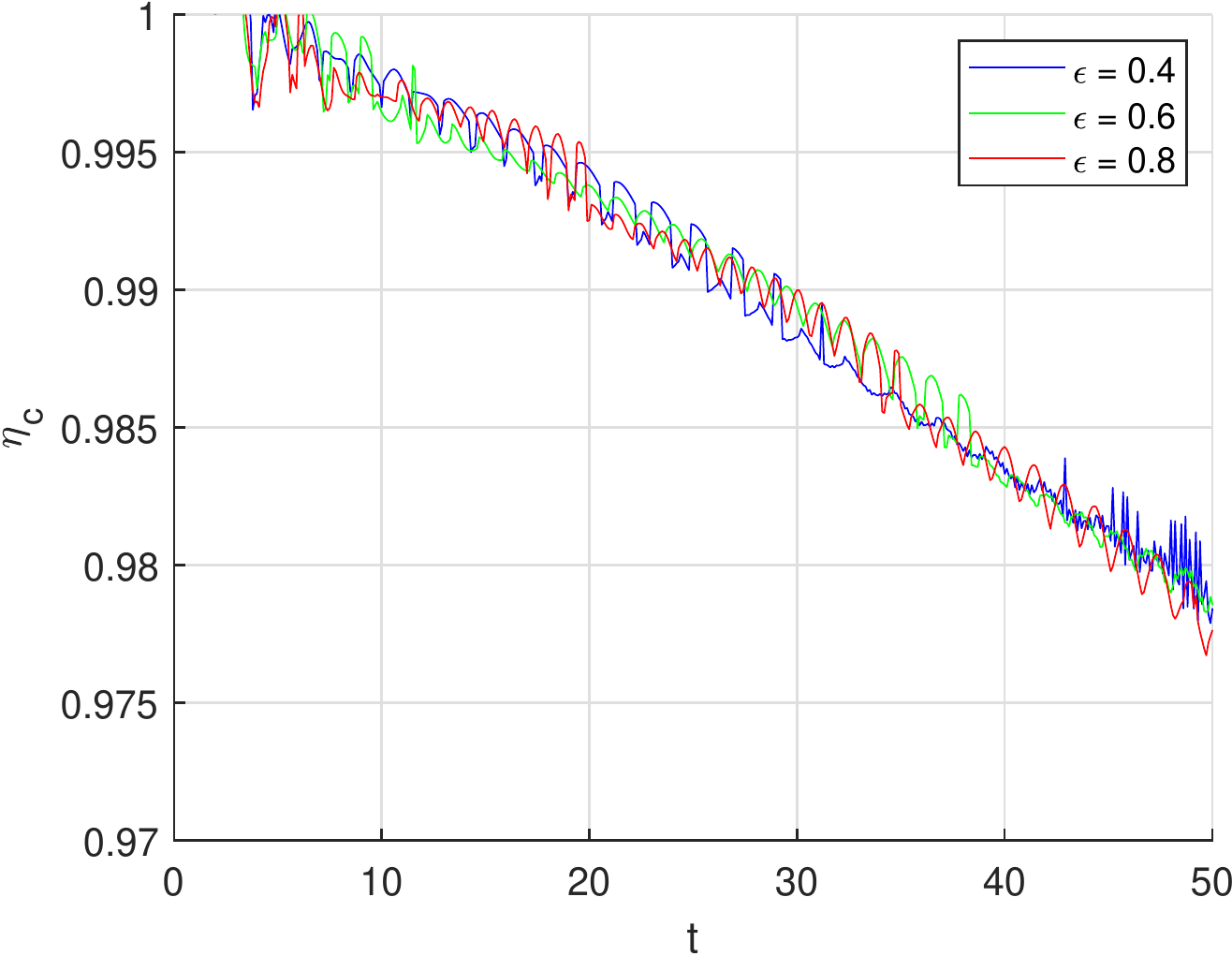}
	\caption{}
	\end{subfigure}
	\caption{(a) The normalised value of $\psi_c$ as a function of $t$, the solid line shows our numerical results, the dashed line gives our analytical prediction from solving \cref{eq:energy_asymp,eq:peak_vor} and the dot-dashed lines give our asymptotic prediction from \cref{eq:U_a_evol_limit}. (b) The normalised value of $\eta_c$ from our numerical simulation as a function of time. Results in both panels are shown for $(\epsilon,U_0,a_0) = (0.4,0.98,1)$ (blue), $(\epsilon,U_0,a_0) = (0.6,1.1,0.9)$ (green) and $(\epsilon,U_0,a_0) = (0.8,1.15,1)$ (red) with $\beta = 0.1$, $D = 25.6$.}
    \label{fig:fig7}
\end{figure}

\cref{fig:fig6} shows the streamfunction, $\psi$, from our numerical simulation with $(\epsilon,\beta,D,U_0,a_0) = (0.8,0.1,25.6,1.15,1)$. For these parameters we expect two shelf wave modes with alongshore wavelengths of $\lambda_n = 26.7$ and $\lambda_n = 42.7$. We observe evidence of these two modes and note that the mode with the shorter alongshore wavelength (and hence larger alongshore wavenumber $k_n$ and smaller offshore wavenumber $l_n$), propagates slower in the $x$ direction due to a more negative value of the group velocity, $c_g-U_f$, and can be seen looping around the domain ahead of the slower mode.

\cref{fig:fig7} shows the values of $\psi_c$ and $\eta_c$ from three simulations in which our theoretical predictions are seen to be accurate. We plot two predictions; firstly, the numerical solution to \cref{eq:energy_asymp,eq:peak_vor} using a fourth order Runge-Kutta scheme is shown with dashed lines and, secondly, the polynomial approximation from \cref{eq:U_a_evol_limit} is shown with dot-dashed lines. In \cref{fig:fig7}.(a), close agreement is observed between our numerical results and both predictions and due to the sensitive dependence of these predictions on $U_0$ and $a_0$ it is difficult to determine which is closest to the numerical results. The accuracy of \cref{eq:U_a_evol_limit} is particularly interesting given that while the condition of $4\epsilon/U \gg \beta$ holds for the parameters we consider, the number of modes, $N$, is not large. The numerical value of $\psi_c$ appears to slowly oscillate relative to the theoretical prediction; this is likely a consequence of some higher order wavelike behaviour within the vortex.

In \cref{fig:fig7}.(b) we plot the value of the maximum vortex vorticity, $\eta_c$, as a function of time in order to test our assumption (see \cref{eq:centre_vor}) that this quantity is conserved over the decay scale of the vortex. We observe that $\eta_c$ decreases by around $2\%$ over the course of the simulations and since this is much smaller than the decrease in $\psi_c$ and similar in magnitude to the effects of viscous dissipation we believe that this assumption is valid.

A supplementary movie file (Movie.mp4) is included to show the temporal evolution of the wave field from the simulation with $(\epsilon,\beta,D,U_0,a_0) = (0.8,0.1,25.6,1.15,1)$ shown in \cref{fig:fig6,fig:fig7}. We plot the mass fluxes, $(Hu,Hv)$, as functions of time, $t$, and position, $(x,y)$, in a frame moving with constant speed, $U_f = 1$. The formation of two modes with different structure and speed can be clearly seen. Additionally, we observe that the waves looping around the domain due to the periodic $x$ direction are unlikely to have a significant effect on the vortex for $t \lesssim 50$. Further, the use of rigid boundary at $y = 51.2$ rather than the semi-infinite domain considered in our theoretical model is justified as there is no noticeable disturbance near $y = 51.2$.

\section{The steady nonlinear problem}
\label{sec:nonlin_steady}

We have shown that if the speed of the vortex matches the topographic wave speed for some wavenumber then the vortex will generate waves and decay due to the transfer of energy from the vortex to the wave field. If however there is no wave speed which matches the vortex speed, we expect steady vortex solutions to exist. This can occur if the vortex moves in the opposite direction to the waves or moves faster than the fastest topographic wave. The conditions for a decaying vortex are given in \cref{eq:cond_wave} and we will focus here on how to determine steady vortex solutions to the full nonlinear system when these conditions are not satisfied.

Neglecting the time derivative in \cref{eq:QG} and combining the constant advection term with the Jacobian gives
\begin{equation}
\label{eq:jac_eqn}
J\left[\psi+U\!\int\! H \dint y,\frac{\zeta+\epsilon}{H}\right] = 0,
\end{equation}
hence we have that the potential vorticity can be written as a function of the total streamfunction as
\begin{equation}
\frac{\zeta+\epsilon}{H} = F\left(\psi+U\!\int\! H \dint y\right).
\end{equation}
The function $F$ may now be determined outside the vortex using the far field condition that $\zeta,\,\psi \to 0$ as $x \to \infty$ and hence
\begin{equation}
F\left(U\!\int\! H(y) \dint y\right) = \frac{\epsilon}{H(y)},
\end{equation}
for all $y$. Defining
\begin{equation}
A(y) = \int_0^y H(y') \dint y',
\end{equation}
as the cross-sectional area in the offshore region $[0,y]$ gives that
\begin{equation}
F(z) = \frac{\epsilon}{H\left(A^{-1}\left(z/U\right)\right)}.
\end{equation}
The full nonlinear problem outside the vortex can now be written as
\begin{equation}
\label{eq:nonlin_prob_out}
\frac{1}{H}\pder{^2\psi}{x^2}+\pder{}{y}\left[\frac{1}{H}\pder{\psi}{y}\right] + \epsilon = \frac{\epsilon H(y)}{H\left(A^{-1}\left(\psi/U+A(y)\right)\right)},
\end{equation}
and solved subject to
\begin{equation}
\label{eq:nonlin_prob_bc_out}
\begin{cases}
\psi = 0 &\quad \textrm{on} \quad y = 0,\\
\psi\to 0 &\quad \textrm{as} \quad x^2+y^2 \to \infty,\\
\psi+UA(y) = 0 &\quad \textrm{on} \quad \mathcal{C},\\
(u,v)\cdot \hat{\textbf{t}} = u_t & \quad \textrm{on} \quad \mathcal{C},
\end{cases}
\end{equation}
where $\mathcal{C}$ is the vortex boundary and $\hat{\textbf{t}}$ is the tangent vector on $\mathcal{C}$. Here $u_t$ is the tangential velocity inside the vortex which is unknown at this stage. The third boundary condition is the no-normal flow condition which states that the vortex boundary is a streamline of the total streamfunction $\Psi = \psi+UA(y)$. Note that this solution is only valid outside the vortex since there are no streamlines which leave the vortex. Therefore inside the vortex we must instead impose $F$. Inside the vortex, $\psi$ satisfies
\begin{equation}
\label{eq:nonlin_prob_in}
\frac{1}{H}\pder{^2\psi}{x^2}+\pder{}{y}\left[\frac{1}{H}\pder{\psi}{y}\right] + \epsilon = H F(\psi+UA(y)),
\end{equation}
subject to
\begin{equation}
\label{eq:nonlin_prob_bc_in}
\begin{cases}
\psi = 0 &\quad\textrm{on}\quad y = 0,\\
\psi+UA(y) = 0 &\quad\textrm{on}\quad \mathcal{C},\\
(u,v)\cdot \hat{\textbf{t}} = u_t & \quad \textrm{on} \quad \mathcal{C}.
\end{cases}
\end{equation}
To obtain a full solution we need to determine the vortex boundary, $\mathcal{C}$ and the function, $F$. This can be achieved by fixing $F$ and then determining $\mathcal{C}$ from the requirement that $\psi+UA = 0$ on $\mathcal{C}$ and $u_t$ is the continuous across $\mathcal{C}$. The typical analytical approach would be instead to impose a boundary and then use $u_t$ from the exterior solution to set a parameter in $F$ \citep{MeleshkoH94,MOFFATT69}, however due to the complicated functional dependence on $H$ and nonlinear nature of this problem, there is unlikely to exist a simple expression for the boundary.

To proceed we take
\begin{equation}
F(z) = -K^2 z + \epsilon,
\end{equation}
inside the vortex. In the case of $H = \textrm{const.}$, the system can be solved exactly to obtain the circular Lamb-Chaplygin dipole \citep{MeleshkoH94} vortex solution centered at a point on the wall. For the Lamb-Chaplygin dipole we have $K = j_1/a$ where $j_1 \approx 3.83$ is the first non-zero root of the Bessel function, $J_1(z)$, and $a$ is the vortex radius. For the case of arbitrary $\epsilon$ and $H$ we expect $K = K(\epsilon,\left< H\right>,a)$ where $\left< H\right>$ denotes some list of parameters of $H$ and does not depend on $y$. Therefore imposing a value for $K$ sets the size of the vortex as a function of $(\epsilon,\left< H\right>)$. We note that the vortex boundary, $\mathcal{C}$, will not necessarily be a semi-circle as in the Lamb-Chaplygin dipole case hence $a$ here is a parameter describing the vortex size rather than its radius. We do, however, expect that the vortex will be close to a circle if $\epsilon\lesssim 1$ and $H(y)$ varies slowly inside the vortex.

We can now seek numerical solutions by choosing a value for $K$ and solving \cref{eq:nonlin_prob_out,eq:nonlin_prob_in} to obtain the streamfunction, $\psi$, and hence the vortex size and velocity fields. The numerical method is described in \cref{sec:app}.

\subsection{Results}

To illustrate our results we consider $H$ given by \cref{eq:H_def}. The cross-sectional area, $A$, is now given by
\begin{equation}
A(y) = \begin{cases}
\frac{1}{\beta}\left[\exp[\beta y] - 1 \right], & y \leq D,\\
\frac{1}{\beta}\left[\exp[\beta D] - 1 \right]+(y-D)\exp[\beta D], & y \geq D,
\end{cases}
\end{equation}
which can be easily inverted for $A^{-1}(z)$.

We will consider here the case where the vortex radius is less than the shelf width, $a < D$, however our theory and numerical method is also valid for $a > D$. Here, the linear operator, $\mathcal{L}$, from \cref{eq:phi_eqn_nonlin} can be written as
\begin{equation}
\mathcal{L} = \nabla^2 +  K^2 \exp[2\beta y]\, \theta(a-r) + \left(\frac{\epsilon \beta}{U}\theta(r-a)-\frac{\beta^2}{4} \right)\theta(D-y),
\end{equation}
which reduces to the linear operator for the Lamb-Chaplygin dipole problem
\begin{equation}
\mathcal{L} = \nabla^2 + K^2 \theta(a-r),
\end{equation}
in the case of small $\epsilon,\,\beta$. The terms $C$ and $N(\phi)$ from \cref{eq:phi_eqn_nonlin} can be similarly expressed in terms of $\beta$ and $D$.

\begin{figure}
	\centering
	\begin{subfigure}[b]{0.49\textwidth}
	\centering
	\includegraphics[trim={0 0 0 0},clip,width=\textwidth]{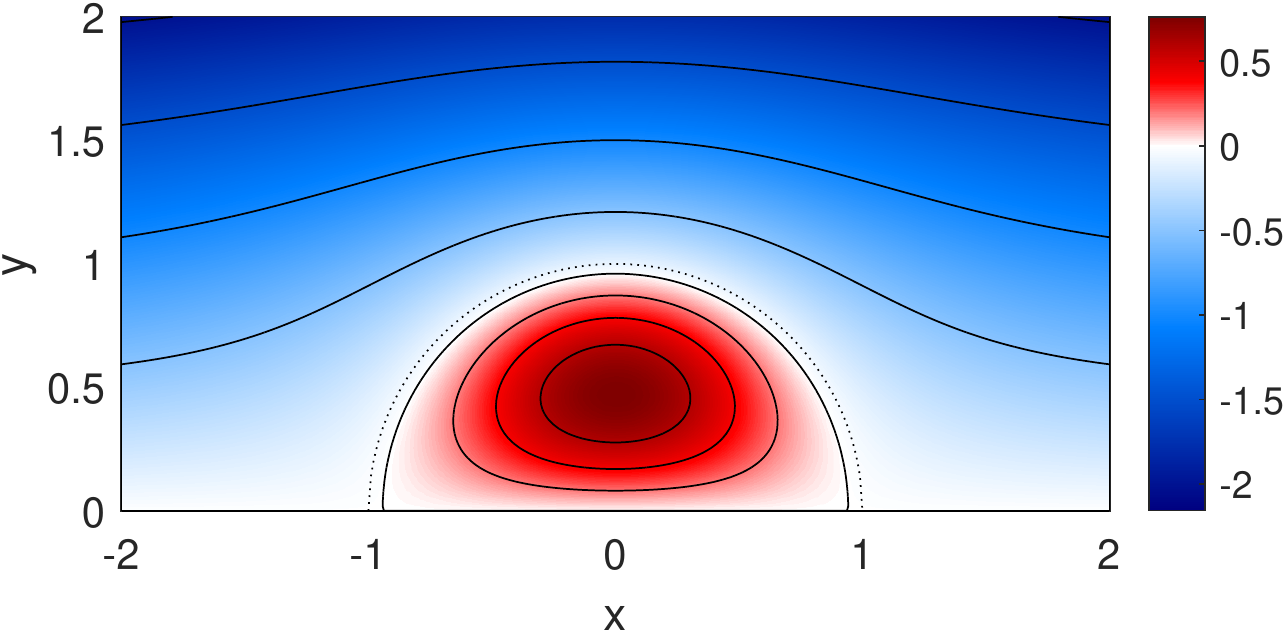}
	\caption{}
	\end{subfigure}
	\begin{subfigure}[b]{0.49\textwidth}
	\centering
	\includegraphics[trim={0 0 0 0},clip,width=\textwidth]{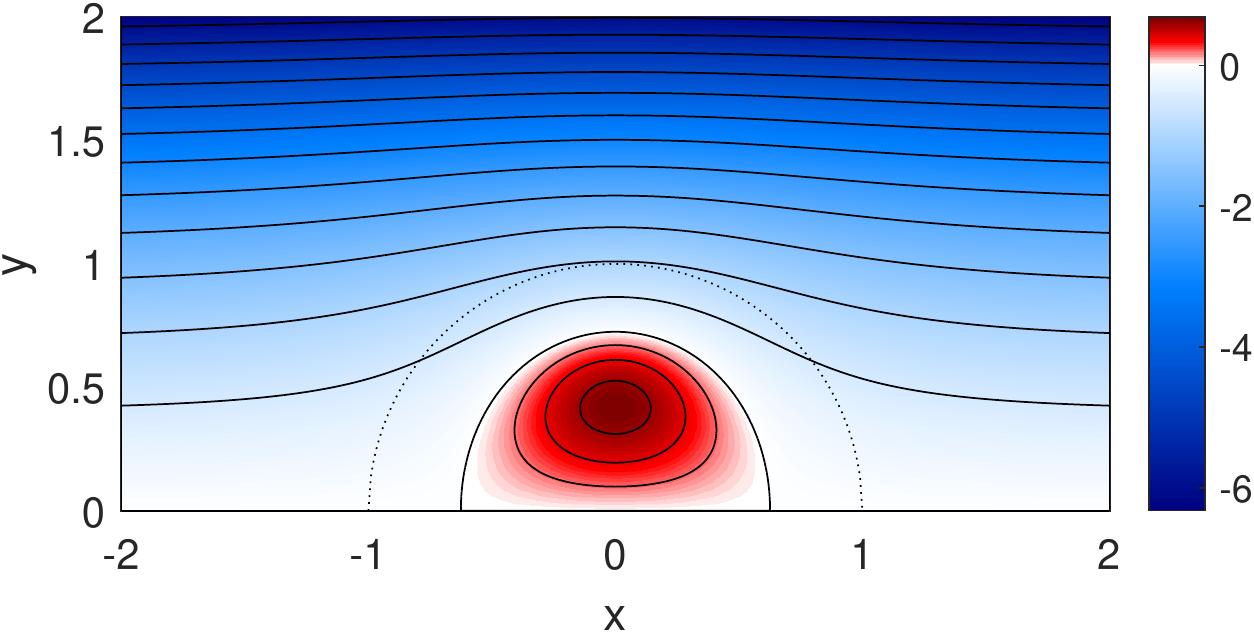}
	\caption{}
	\end{subfigure}
	\newline
	\begin{subfigure}[b]{0.49\textwidth}
	\centering
	\includegraphics[trim={0 0 0 0},clip,width=\textwidth]{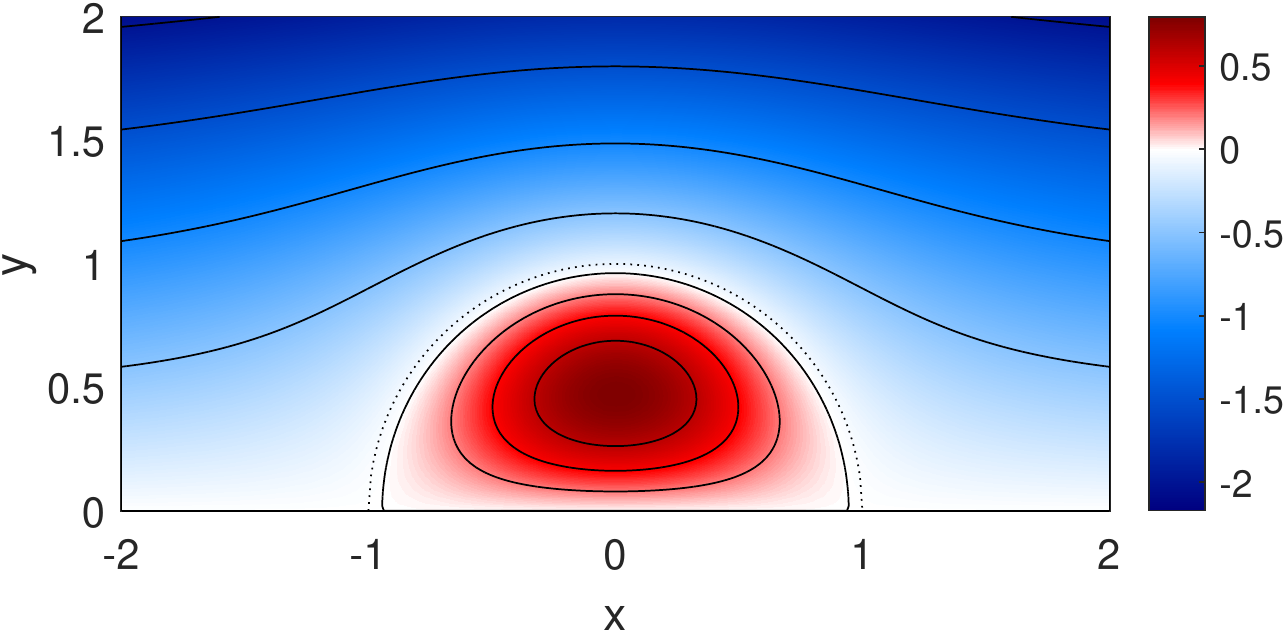}
	\caption{}
	\end{subfigure}
	\begin{subfigure}[b]{0.49\textwidth}
	\centering
	\includegraphics[trim={0 0 0 0},clip,width=\textwidth]{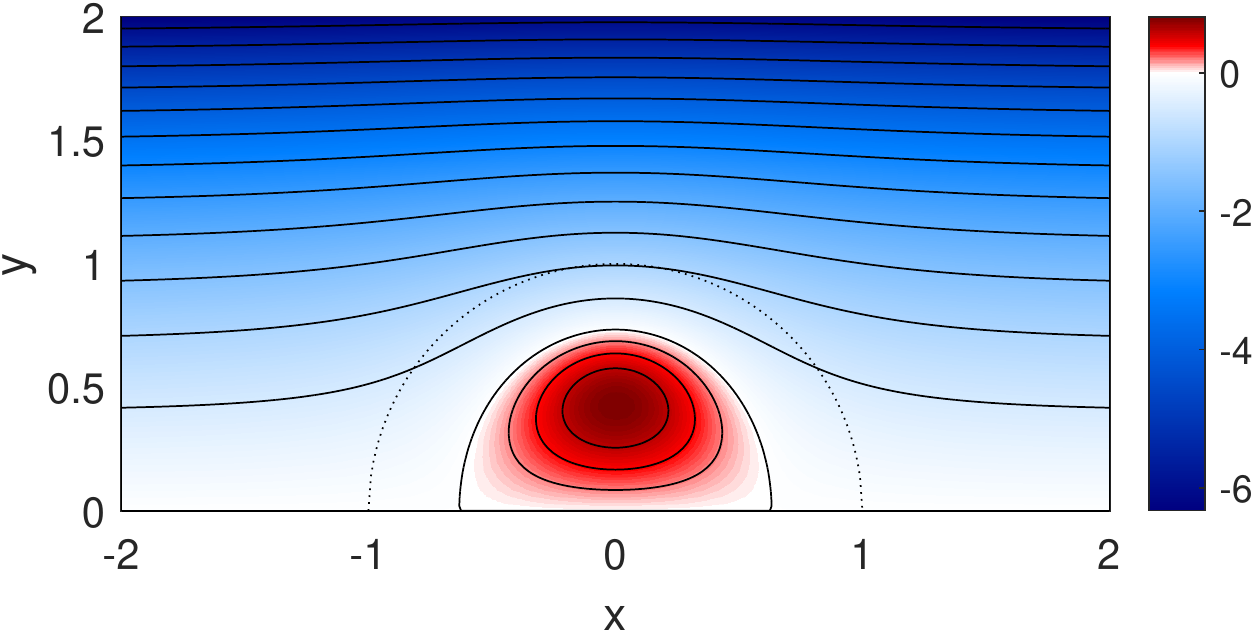}
	\caption{}
	\end{subfigure}
	\newline
	\begin{subfigure}[b]{0.49\textwidth}
	\centering
	\includegraphics[trim={0 0 0 0},clip,width=\textwidth]{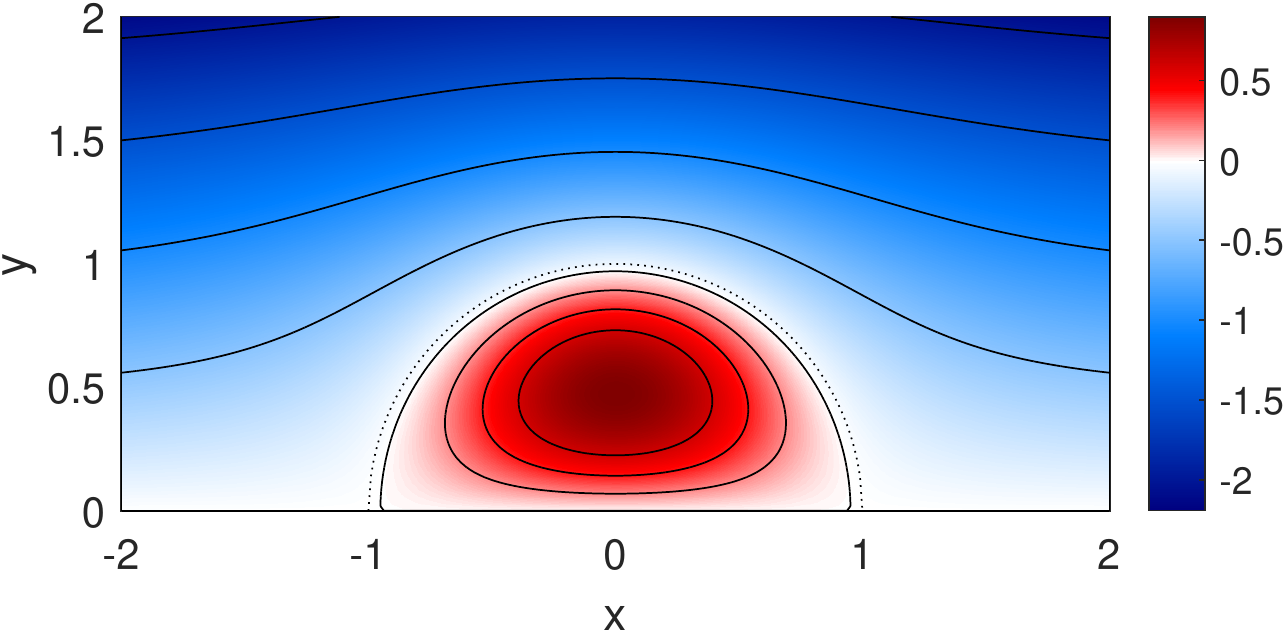}
	\caption{}
	\end{subfigure}
	\begin{subfigure}[b]{0.49\textwidth}
	\centering
	\includegraphics[trim={0 0 0 0},clip,width=\textwidth]{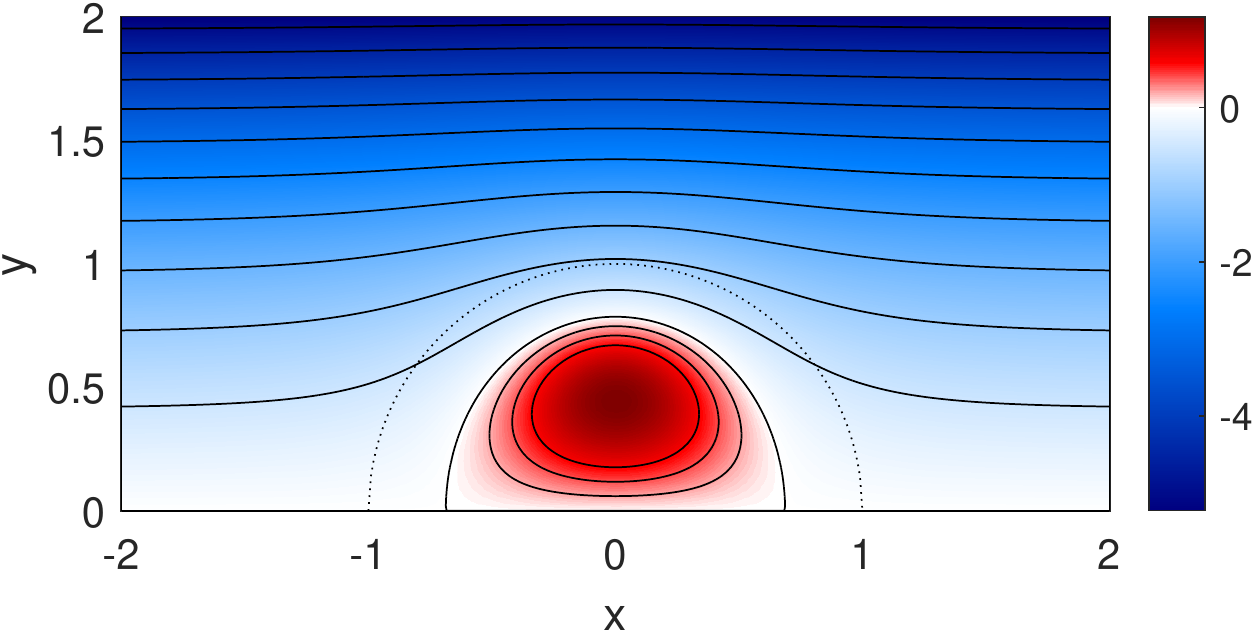}
	\caption{}
	\end{subfigure}
	\caption{Plots of the streamfunction in the vortex frame, $\psi + UA(y)$, for 6 pairs of parameters, $(\epsilon,\beta)$, given by (a) $(0.25,0.1)$, (b) $(0.25,1)$, (c) $(1,0.1)$, (d) $(1,1)$, (e) $(4,0.1)$, and (f) $(4,1)$. Solutions are shown for $K = j_1$, $U = -1$ and $D = 12.5$. Dependence on $D$ is weak and hence not shown here. The dotted line shows the vortex boundary for the Lamb-Chaplygin dipole case of $(\epsilon,\beta) = (0,0)$ using the same value of $K$.}
    \label{fig:fig8}
\end{figure}

\begin{figure}
	\centering
	\begin{subfigure}[b]{0.49\textwidth}
	\centering
	\includegraphics[trim={0 0 0 0},clip,width=\textwidth]{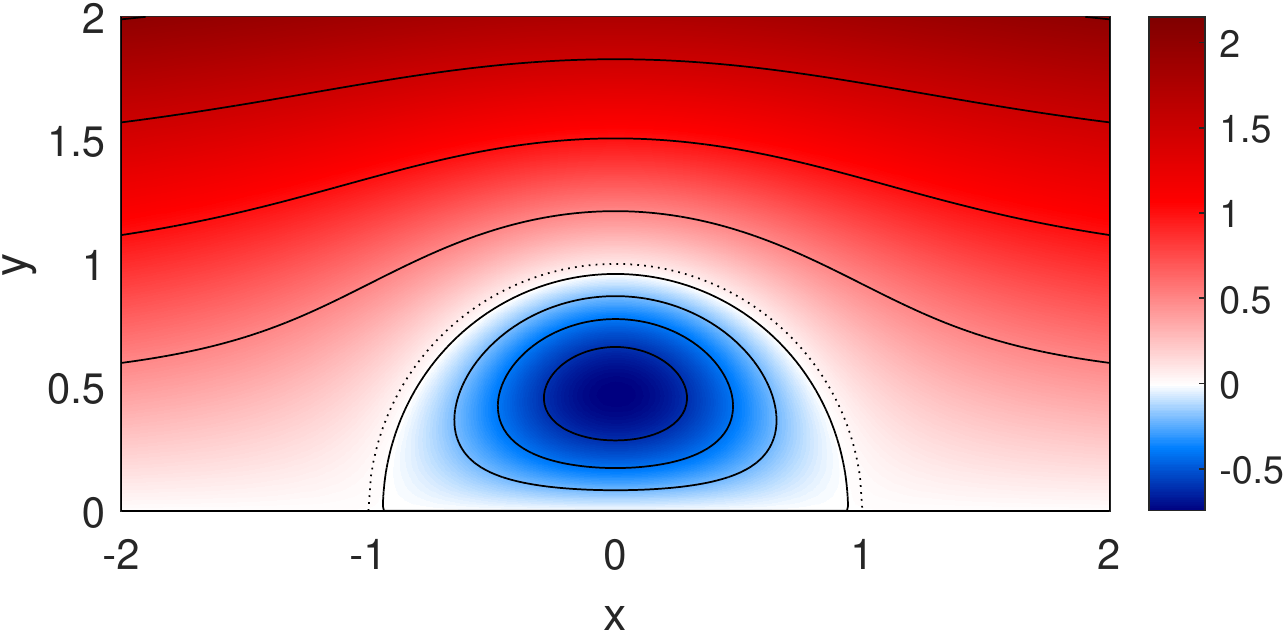}
	\caption{}
	\end{subfigure}
	\begin{subfigure}[b]{0.49\textwidth}
	\centering
	\includegraphics[trim={0 0 0 0},clip,width=\textwidth]{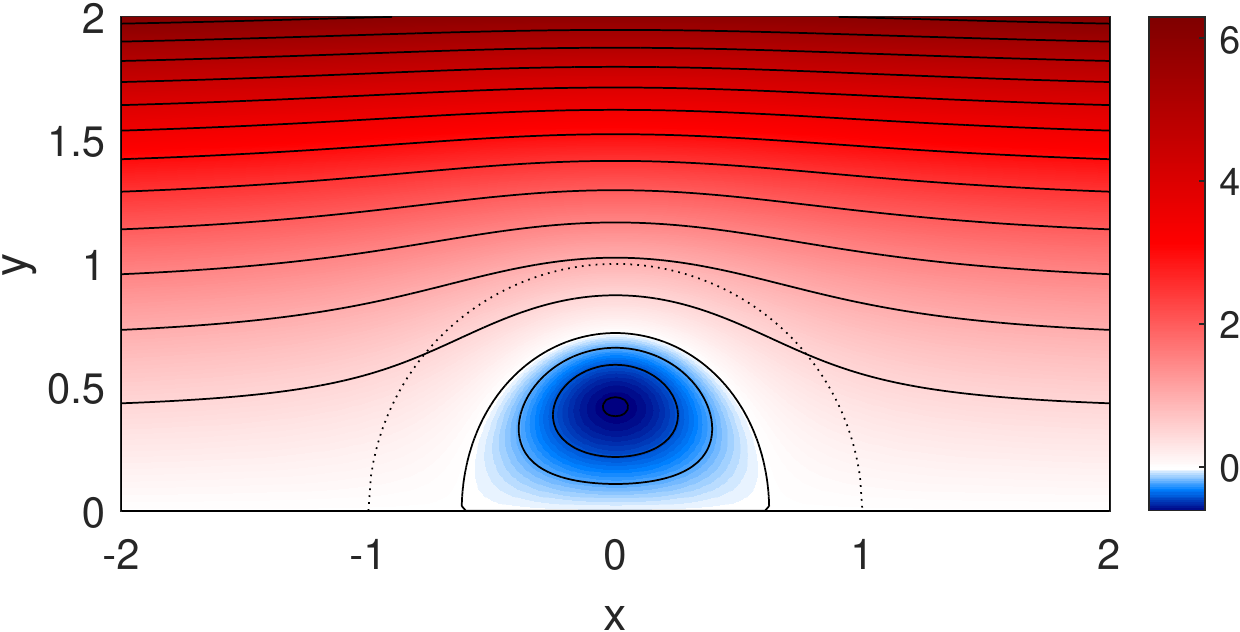}
	\caption{}
	\end{subfigure}
	\caption{Plots of the streamfunction in the vortex frame, $\psi + UA(y)$, for $\epsilon = 0.1$, $U = 1$ and $\beta = 0.1$ (a) and $\beta = 1$ (b). Solutions are shown for $K = j_1$ and $D = 12.5$ and the dotted line shows the vortex boundary for the Lamb-Chaplygin dipole case of $(\epsilon,\beta) = (0,0)$ using the same value of $K$. These solution correspond to vortices travelling faster than the fastest topographic mode. While the dependence of the vortex structure on $D$ is weak, the value of $D = 12.5$ does ensure that there are no modes matching the vortex speed.}
    \label{fig:fig9}
\end{figure}

\cref{fig:fig8,fig:fig9} shows our nonlinear solutions for $\psi$ for a range of parameters. \cref{fig:fig8} shows vortices which travel in the opposite direction to the wave field ($U = -1$) for parameter values $(\epsilon,\beta) = \{0.25, 1, 4\}\times\{0.1, 1\}$ while \cref{fig:fig9} shows vortices which travel faster than the fastest topographic mode for $\epsilon = 0.1$ and $\beta \in \{0.1, 1\}$. The solutions are calculated on the numerical domain $(x,y) \in [-51.2,\, 51.2]\times[0,\, 51.2]$ using $(Nx,Ny) = (2048,1024)$ grid-points and the solution is assumed to have converged if $\delta = 10^{-10}$ in \cref{eq:convergence_cond}. We plot the streamfunction in the frame of the vortex, $\psi+UA(y)$, so the streamline of height $0$ denotes the vortex boundary. While our system depends on 5 parameters, $\epsilon$, $\beta$, $D$, $U$ and $K$, we can set $|U| = 1$ and fix $K$. Setting $|U| = 1$ is equivalent to setting the velocity scale used for calculating the inverse Rossby number, $\epsilon$, whereas fixing $K$ determines the size of the vortex and hence we can measure the slope, $\beta$, and shelf width, $D$, in units of the Lamb-Chaplygin dipole radius, $a = j_1/K$. We therefore vary only $\epsilon$, $\beta$ and $D$ and note that the dependence of the vortex structure on  $D$ is weak and not shown here. We note, however, that $D$ can play an important role in setting the speed of the modes and hence is chosen such that the vortices in \cref{fig:fig9} do not generate topographic modes. For the decaying vortex problem where energy is emitted towards the edge of the shelf we expect that $D$ will play a more important role.

From \cref{fig:fig8,fig:fig9} we observe that the effect of increasing $\beta$ is to reduce the vortex size and to slightly alter its aspect ratio. Conversely, increasing $\epsilon$ has no significant effect on the vortex size and shape though it does increase the peak value of the streamfunction, corresponding to an increase in peak vorticity. The vortex shape can be discussed in terms of aspect ratio; the ratio of the offshore radius, $a_y$, to the alongshore radius, $a_x$, given by $a_r = a_y/a_x$. Here $a_x$ and $a_y$ are defined as the distance from the origin to the curve $\psi+UA(y) = 0$ in the $x$ and $y$ directions respectively. From \cref{eq:jac_eqn} we note that $\psi$ and $\zeta$ may be scaled on $U$ and hence $\epsilon$ enters the system only through the quantity $\epsilon/U$. Therefore, on dimensional grounds we may write the maximum vorticity as
\begin{equation}
\zeta_{max} = \frac{U}{a} G\left(\frac{\epsilon}{U},\beta,D\right),
\end{equation}
where $G$ is some function describing the dependence of the maximum vorticity on the remaining parameters and $a$ is some parameter describing the vortex size, taken here as the offshore radius, $a = a_y$.

\begin{figure}
	\centering
	\begin{subfigure}[b]{0.49\textwidth}
	\centering
	\includegraphics[trim={0 0 0 0},clip,width=\textwidth]{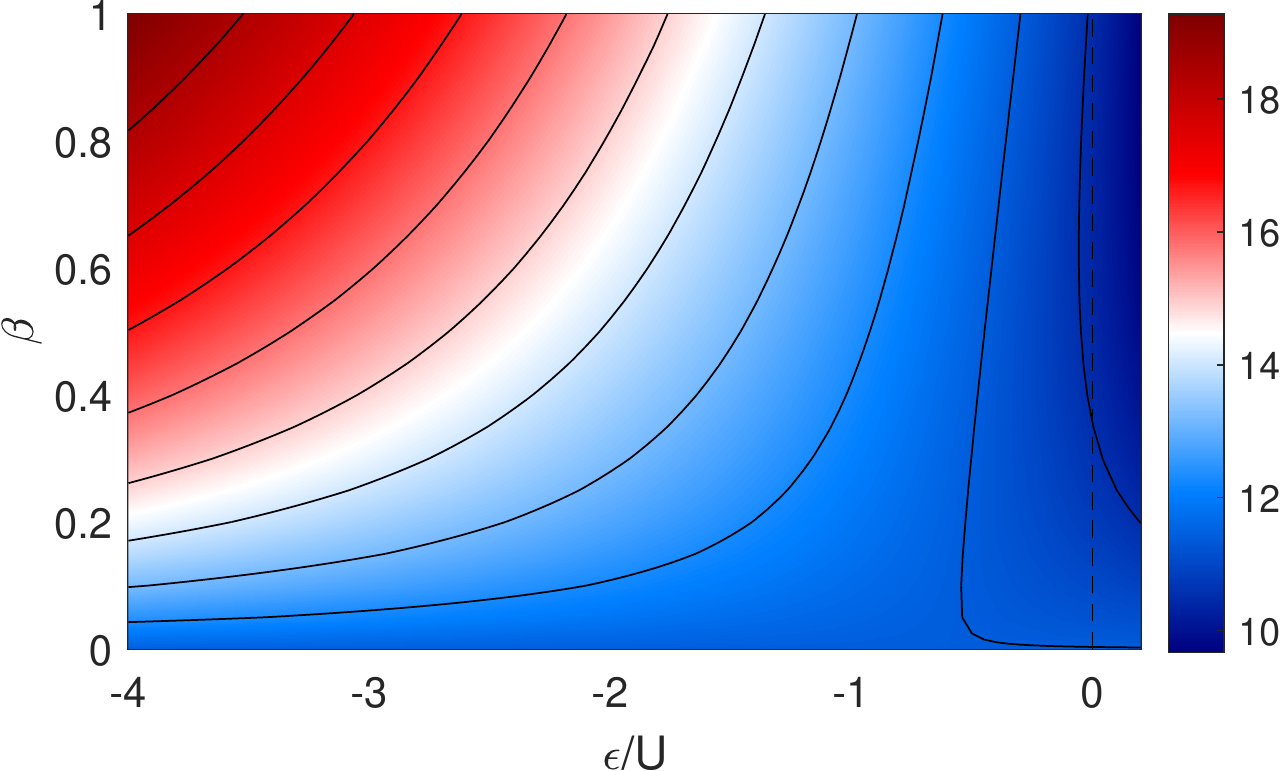}
	\caption{}
	\end{subfigure}
	\begin{subfigure}[b]{0.49\textwidth}
	\centering
	\includegraphics[trim={0 0 0 0},clip,width=\textwidth]{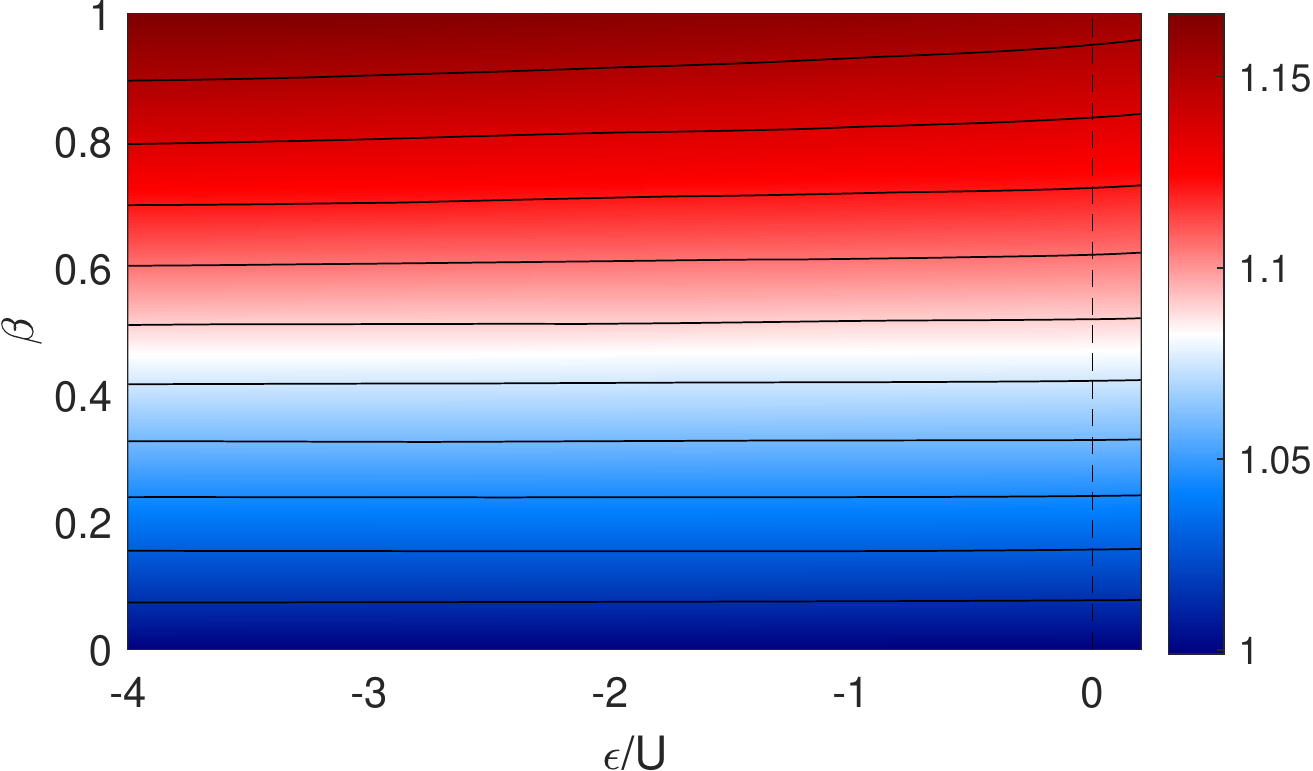}
	\caption{}
	\end{subfigure}
	\caption{Plots of $G$ (a) and $a_r$ (b) as functions of $\epsilon/U$ and $\beta$ for $D = 12.5$. The range of parameters shown is limited for $\epsilon/U > 0$ due to the appearance of decaying vortices in this region (we plot results for $\epsilon/U \in [-4, 0.2]$). The dashed line denotes $\epsilon/U = 0$.}
    \label{fig:fig10}
\end{figure}

In \cref{fig:fig10} we plot $G$ and $a_r$ as functions of $\epsilon/U$ and $\beta$ for $D = 12.5$. Dependence on $D$ is weak and not discussed here. Wave generation occurs for positive values of $\epsilon/U$ (see \cref{eq:cond_wave}) and we find that our iterative method fails to converge for vortices close to the decaying regime. Therefore only a small region of parameter space is shown for small $\epsilon/U > 0$ corresponding to vortices moving faster than all wave modes as in \cref{fig:fig9}. The value of $G$ is seen to increase with both $\beta$ and $-\epsilon/U$ corresponding to a higher vorticity within the vortex centre. Finally, we observe that the aspect ratio, $a_r$, increases slowly with $\beta$ and has very weak dependence on $\epsilon/U$. In the case of $\beta = 0$, the results are independent of $\epsilon$ and correspond to the Lamb-Chaplygin case given in \cref{eq:LCD}.

\section{Discussion and conclusions}
\label{sec:diss}

Here we have considered the evolution of a vortex moving along a shelf. To allow for the calculation of analytical results in the limit of shallow slope and slow rotation, we take our vortex to be contained within an approximately semi-circular region against the wall. Vortices of this form therefore limit to one half of the Lamb-Chaplygin dipole with the other half corresponding to the vortex image.

Since a shelf system admits shelf wave solutions, we began by determining the speed and structure of these modes. For positive rotation rate, these shelf waves move with the coastal boundary on the right as expected for coastal trapped waves. The finite shelf width acts to discretise the modes leading to a countable set of wave solutions. The alongshore and offshore wavenumbers can be determined numerically by solving a transcendental equation and the frequency and phase speed of each mode can then be determined.

If the speed of the vortex matches the phase speed of any shelf wave modes, we expect the vortex to excite these modes generating a wave wake. Using a Fourier transform approach, we have determined the far-field amplitude of the wave wake and provided analytic predictions for the number of modes generated as a function of the vortex speed, shelf parameters and rotation rate. We observe that a slower vortex will match the phase speeds of a greater number of waves and will hence generate a higher number of different modes. The group velocity of each modes is less than or equal to its phase velocity hence all modes will be emitted behind the vortex and we expect no upstream wave signature.

The generation of waves corresponds to a flux of energy from the vortex to the wave field with this flux resulting in the slow decay of the vortex. We have determined the leading order wave energy flux using our far field wave solution and by equating this flux to the loss of vortex energy we can describe the vortex decay. This decay is shown to be proportional to the square of the vortex dipole strength and to have a complicated dependence on the vortex speed, rotation rate and shelf slope. The vortex slows as it loses energy, and so excites additional modes with large alongshore wavelength. The appearance of new modes leads to a peak in energy flux resulting more rapid loss of energy. As the vortex slows further, the energy flux from this new mode and the vortex decay rate decrease until a new mode appears.

In the limit of small rotation rate and small shelf slope we can approximate our vortex to leading order using the Lamb-Chaplygin solution. This gives  analytical expressions for the vortex energy and dipole strength which allows us to solve for the evolution of the vortex speed and radius by numerically integrating the energy balance equation. Additionally, we present approximate analytic solutions for the case where the number of modes is large, corresponding to either a very slow vortex or a wide shelf region. Polynomial decay of the vortex speed and radius are observed and the results are consistent with the infinite width shelf limit which can be derived using the methods of \citet{JohnsonCrowe20} and \citet{Croweetal20}.

To test our predictions we present numerical simulations of the full nonlinear system. The vortex generated wave fields are shown for a range of parameters and found to be consistent with our predictions. Finally, we compare our predicted vortex decay with the vortex decay observed from numerical simulations and demonstrate fairly close agreement. We note, however, that for a general rotation rate and shelf slope the difficulties in accurately determining the vortex energy and dipole strength make it hard to test our predictions due to the sensitivity of our results on these quantities. Additionally, the energy flux can rapidly increase as a new mode appears, therefore any inaccuracies in estimating the vortex speed can lead to significant errors in the vortex decay rate when the vortex is close to exciting a new mode.

If the vortex does not excite any wave modes - either by travelling faster than the fastest wave or travelling in the opposite direction to all shelf waves - we expect steady vortex solutions to exist. We therefore consider the full nonlinear problem and present a numerical approach for determining these nonlinear vortex solutions. Finally, solutions are presented for a range of rotation rates and shelf slopes and compared to the Lamb-Chaplygin limit. We find that increasing the rotation rate, hence reducing the Rossby number, increases the maximum vorticity of the vortex. Increasing the shelf slope also corresponds to an increase of the maximum vorticity when compared to the Lamb-Chaplygin case. In addition, increasing shelf slope also changes the shape of the vortex boundary with a slightly increased offshore scale compared to the alongshore scale.

From \cref{eq:U_a_evol_limit} we may estimate the dimensional decay timescale of a vortex as
\begin{equation}
T \sim \sqrt{\frac{U}{a \beta^3 f^3}},
\end{equation}
where here $U$ and $a$ describe the dimensional speed and radius of the vortex and $\beta$ measures the fractional change in depth across the vortex. Note that this result is valid for $4 a f/U \gg \beta$ which ensures that there are many shelf wave modes matching the vortex speed. Typical values of $U = 0.1 \textrm{ms}^{-1}$, $a = 4\times10^2 \textrm{m}$, $f = 10^{-4} \textrm{s}^{-1}$ and $\beta = 0.1$ gives a timescale of around $5$ days. Our asymptotic model requires a small inverse Rossby number so is only valid for `small scale' structures where rotation is dominated by advection.

We have used an exponential shelf profile throughout to illustrate our method and results. For shelf waves above general monotonically sloping topography, \citet{Huthnance74} and \citet{GillSchumann74} show that the dispersion relation takes a similar form to that here and the inner product required for the orthogonality of modes exists. We thus expect our results to extend to such shelf profiles and, when wave mode phase speeds match the vortex speed, we expect vortex decay at a rate dependent on the shelf slope in the same manner as for exponential topography.

The use of a shallow water model with a rigid lid assumption results in several limitations. Firstly, the effects of vertical stratification, which may be expected to be important over the scales of coastal vortices, are ignored. This precludes the consideration of baroclinic effects such as depth dependence in the vortex and the generation of internal Kelvin waves \citep{DEWARHOGG,DeMarezetal17,DeMarezetal20}. Secondly, the rigid lid approximation eliminates the free surface Kelvin and Poincar\'e waves. The relevant parameter determining the strength of these waves is the Froude number
\begin{equation}
F = \frac{U}{\sqrt{gH}},
\end{equation}
which is typically small for coastal systems \citep{GillSchumann74}. The vortex speed $U$ is small compared to the longwave speed $\sqrt{gH}$ and  free-surface surface modes are only weakly generated. The ratio of wave energy lost to surface waves \citep{Fordetal2000} to that lost to shelf waves is of order $O(F^3/\epsilon^3)$ which is small for the parameters considered above.\\

%\noindent{\bf  Supplementary material\bf{.}} The Dedalus setup file and a supplementary movie are available at ...\\

\noindent{\bf Funding\bf{.}} This work was funded by the UK Natural Environment Research Council under grant number NE/S009922/1.\\

\noindent{\bf Declaration of Interests\bf{.}} The authors report no conflict of interest.

\appendix
\section{General numerical solution}
\label{sec:app}

Here we present the numerical procedure used for solving \cref{eq:nonlin_prob_in,eq:nonlin_prob_out} subject to the boundary conditions \cref{eq:nonlin_prob_bc_in,eq:nonlin_prob_bc_out}. We begin by noting that
\begin{equation}
\sgn[\psi + UA] = -\sgn[U],
\end{equation}
inside the vortex and
\begin{equation}
\sgn[\psi + UA] = \sgn[U],
\end{equation}
outside to combine \cref{eq:nonlin_prob_out,eq:nonlin_prob_in} as
\begin{multline}
\label{eq:nonlin_prob_comb}
\nabla^2\psi-\frac{H_y}{H}\pder{\psi}{y} + \epsilon H = \frac{\epsilon H^2}{H\left(A^{-1}\left(\psi/U+A\right)\right)}\theta(\psi/U+A) \\ +H^2\left[ \epsilon-K^2\left(\psi+UA\right)\right]\theta(-\psi/U-A),
\end{multline}
where $\theta$ is the heaviside function. Substituting $\psi = \sqrt{H}\phi$ gives
\begin{multline}
\label{eq:nonlin_prob_phi}
\nabla^2\phi+\frac{1}{2}\left[\frac{H_{yy}}{H}-\frac{3H_y^2}{2H^2}\right]\phi = -\epsilon \sqrt{H} + \frac{\epsilon H^{3/2}}{H\left(A^{-1}\left(\sqrt{H}\phi/U+A\right)\right)}\theta(\sqrt{H}\phi/U+A) \\ +H^{3/2} \left[\epsilon - K^2\left(\sqrt{H}\phi+UA\right)\right]\theta(-\sqrt{H}\phi/U-A),
\end{multline}
and noting that
\begin{equation}
\frac{\epsilon H^{3/2}}{H\left(A^{-1}\left(\sqrt{H}\phi/U+A\right)\right)} = \epsilon\sqrt{H} - \frac{\epsilon H_y \phi}{U H} + O(\phi^2),
\end{equation}
we may split \cref{eq:nonlin_prob_phi} into linear and nonlinear parts as
\begin{multline}
\label{eq:split_eqn}
\left[ \nabla^2 + \left( K^2 H^2 \theta(a-r) + \frac{\epsilon H_y}{UH}\theta(r-a)+\frac{1}{2}\left[\frac{H_{yy}}{H}-\frac{3H_y^2}{2H^2}\right] \right)\right]\phi = \\ -\sqrt{H}\left[\epsilon(1-H)+K^2 H U A\right]\theta(a-r) + N(\phi),
\end{multline}
where
\begin{multline}
N(\phi) = H^{3/2}\left[\theta(-\sqrt{H}\phi/U-A) - \theta(a-r)\right]\left[\epsilon -K^2 \left(\sqrt{H}\phi+ UA\right)\right] + \\ \theta(r-a)\left[-\epsilon\sqrt{H} + \frac{\epsilon H_y \phi}{U H}\right] + \theta(\sqrt{H}\phi/U+A)\frac{\epsilon H^{3/2}}{H\left(A^{-1}\left(\sqrt{H}\phi/U+A\right)\right)}.
\end{multline}
We note that \cref{eq:split_eqn} is an exact rearrangement of \cref{eq:nonlin_prob_phi} for all values of $a$. However, picking $a$ to be close to the average vortex radius will minimise the nonlinear term, $N$, and make finding solutions easier numerically. Defining the left-hand side linear operator as
\begin{equation}
\mathcal{L} = \nabla^2 + \left( K^2 H^2 \theta(a-r) + \frac{\epsilon H_y}{UH}\theta(r-a)+\frac{1}{2}\left[\frac{H_{yy}}{H}-\frac{3H_y^2}{2H^2}\right] \right),
\end{equation}
and the $\phi$ independent term as
\begin{equation}
C = -\sqrt{H}\left[\epsilon(1-H)+K^2 H U A\right]\theta(a-r),
\end{equation}
we may write \cref{eq:split_eqn} as
\begin{equation}
\label{eq:phi_eqn_nonlin}
\mathcal{L} \phi = C + N(\phi).
\end{equation}
We can now solve \cref{eq:phi_eqn_nonlin} using an iterative method. We begin by finding the linear solution, $\phi_0$, satisfying
\begin{equation}
\mathcal{L} \phi_0 = C,
\end{equation}
by numerically inverting $\mathcal{L}$ and imposing boundary conditions of $\phi = 0$ on the boundaries of the numerical domain. Using this linear solution as our initial guess we may now take
\begin{equation}
\phi_n = \mathcal{L}^{-1}\left[C + N(\phi_{n-1})\right],
\end{equation}
and iterate until the domain averaged difference between consecutive $\phi_n$ is small,
\begin{equation}
\label{eq:convergence_cond}
\int_\mathcal{D} |\phi_{n}-\phi_{n-1}| \dint A < \delta,
\end{equation}
for some $\delta$. Picking a value of $a$ close to the size of the vortex minimises the nonlinear term $N$ and leads to faster and more consistent convergence. It is sufficient to pick initial radius $a$ using the Lamb-Chaplygin dipole value of $a = j_1/K$ for small $\beta$ and $\epsilon$. For larger parameter values we can use gradually increase $\epsilon$ or $\beta$ while adjusting $a$ to match the observed vortex size from the previous parameter values. if convergence from the linear solution is slow or fails, we can use a parameter continuation approach by taking the nonlinear solution for slightly smaller $\epsilon$ or $\beta$ as our initial guess, $\phi_0$.

\bibliographystyle{jfm}
\bibliography{bibliography}

\begin{thebibliography}{28}
\expandafter\ifx\csname natexlab\endcsname\relax\def\natexlab#1{#1}\fi

\bibitem[Bretherton(1967)]{bretherton_1967}
{\sc Bretherton, F.~P.} 1967 The time-dependent motion due to a cylinder moving
  in an unbounded rotating or stratified fluid. {\em J. Fluid Mech.\/} {\bf
  28}~(3), 545–570.

\bibitem[Burns {\em et~al.\/}(2020)Burns, Vasil, Oishi, Lecoanet \&
  Brown]{BurnsVOLB20}
{\sc Burns, K.~J., Vasil, G.~M., Oishi, J.~S., Lecoanet, D. \& Brown, B.~P.}
  2020 Dedalus: {A} flexible framework for numerical simulations with spectral
  methods. {\em Phys. Rev. Res.\/} {\bf 2}, 023068.

\bibitem[Crowe \& Johnson(2020)]{crowe_johnson_2020}
{\sc Crowe, M.~N. \& Johnson, E.~R.} 2020 The effects of vertical mixing on
  nonlinear {K}elvin waves. {\em J. Fluid Mech.\/} {\bf 903}, A22.

\bibitem[Crowe {\em et~al.\/}(2021)Crowe, Kemp \& Johnson]{Croweetal20}
{\sc Crowe, M.~N., Kemp, C. J.~D. \& Johnson, E.~R.} 2021 The decay of {H}ill's
  vortex in a rotating flow. {\em J. Fluid Mech.\/} {\bf 919}~(A6).

\bibitem[Deremble {\em et~al.\/}(2017)Deremble, Johnson \& Dewar]{DEREMBLEETAL}
{\sc Deremble, B., Johnson, E.R. \& Dewar, W.K.} 2017 A coupled model of
  interior balanced and boundary flow. {\em Ocean Modelling\/} {\bf 119}, 1 --
  12.

\bibitem[Dewar {\em et~al.\/}(2011)Dewar, Berloff \& Hogg]{DEWARETAL}
{\sc Dewar, W.~K., Berloff, P. \& Hogg, A.~McC.} 2011 Submesoscale generation
  by boundaries. {\em J. Mar. Res.\/} {\bf 69}~(4-5), 501--522.

\bibitem[Dewar \& Hogg(2010)]{DEWARHOGG}
{\sc Dewar, W.~K. \& Hogg, A.~McC.} 2010 Topographic inviscid dissipation of
  balanced flow. {\em Ocean Modelling\/} {\bf 32}~(1), 1 -- 13.

\bibitem[Flierl \& Haines(1994)]{FLIERLHAINES94}
{\sc Flierl, G.~R. \& Haines, K} 1994 The decay of modons due to {R}ossby wave
  radiation. {\em Phys. Fluids\/} {\bf 6}~(10), 3487--3497.

\bibitem[Ford {\em et~al.\/}(2000)Ford, McIntyre \& Norton]{Fordetal2000}
{\sc Ford, R., McIntyre, M.~E. \& Norton, W.~A.} 2000 Balance and the slow
  quasimanifold: {S}ome explicit results. {\em Journal of the Atmospheric
  Sciences\/} {\bf 57}~(9), 1236--1254.

\bibitem[Fraenkel(1956)]{Fraenkel56}
{\sc Fraenkel, L.~E.} 1956 On the flow of rotating fluid past bodies in a pipe.
  {\em Proc. R. Soc. A.\/} {\bf 233}~(1195), 506--526.

\bibitem[Gill \& Schumann(1974)]{GillSchumann74}
{\sc Gill, A.~E. \& Schumann, E.~H.} 1974 The generation of long shelf waves by
  the wind. {\em J . Phys. Oceanog.\/} {\bf 4}, 83--90.

\bibitem[Hill(1894)]{HILL1894}
{\sc Hill, M. J.~M.} 1894 On a spherical vortex. {\em Phil. Trans. R. Soc.
  (A.)\/} {\bf 185}, 213--245.

\bibitem[Hogg {\em et~al.\/}(2011)Hogg, Dewar, Berloff \& Ward]{HOGGETAL}
{\sc Hogg, A.~McC., Dewar, W.~K., Berloff, P. \& Ward, M.~L.} 2011 {Kelvin}
  wave hydraulic control induced by interactions between vortices and
  topography. {\em J. Fluid Mech.\/} {\bf 687}, 194–208.

\bibitem[Huthnance(1974)]{Huthnance74}
{\sc Huthnance, J.~M.} 1974 On trapped waves over a continental shelf. {\em J.
  Fluid Mech.\/} {\bf 69}, 689--704.

\bibitem[Isern-Fontanet {\em et~al.\/}(2006)Isern-Fontanet, García-Ladona \&
  Font]{IsernFontanetetal2006}
{\sc Isern-Fontanet, J., García-Ladona, E. \& Font, J.} 2006 Vortices of the
  {M}editerranean {S}ea: {A}n altimetric perspective. {\em J. Phys.
  Oceanogr.\/} {\bf 36}, 87 -- 103.

\bibitem[Johnson(1979)]{JOHNSON79}
{\sc Johnson, E.~R.} 1979 Finite depth stratified flow over topography on a
  beta-plane. {\em Geophys. Astrophys. Fluid Dyn.\/} {\bf 12}, 35 -- 43.

\bibitem[Johnson(1989)]{johnson89}
{\sc Johnson, E.~R.} 1989 Topographic waves in open domains. {P}art 1.
  {B}oundary conditions and frequency estimates. {\em J. Fluid Mech.\/} {\bf
  200}, 69–76.

\bibitem[Johnson \& Crowe(2021)]{JohnsonCrowe20}
{\sc Johnson, E.~R. \& Crowe, M.~N.} 2021 The decay of a dipolar vortex in a
  weakly dispersive environment. {\em J. Fluid Mech.\/} {\bf (in press)}.

\bibitem[Johnson \& Rodney(2011)]{JOHNSON2011}
{\sc Johnson, E.~R. \& Rodney, J.~T.} 2011 Spectral methods for coastal-trapped
  waves. {\em Continental Shelf Res.\/} {\bf 31}~(14), 1481 -- 1489.

\bibitem[LeBlond \& Mysak(1978)]{LeBlondM78}
{\sc LeBlond, P.~H. \& Mysak, L.~A.} 1978 {\em Waves in the {O}cean\/}.
  Elsevier.

\bibitem[Lighthill(1967)]{lighthill_1967}
{\sc Lighthill, M.~J.} 1967 On waves generated in dispersive systems by
  travelling forcing effects, with applications to the dynamics of rotating
  fluids. {\em J. Fluid Mech.\/} {\bf 27}~(4), 725–752.

\bibitem[Long(1953)]{LONG53}
{\sc Long, R.~R.} 1953 Steady motion around a symmetrical obstacle moving along
  the axis of a rotating liquid. {\em J. Met.\/} {\bf 10}, 197--203.

\bibitem[Machicoane {\em et~al.\/}(2018)Machicoane, Labarre, Voisin, Moisy \&
  Cortet]{Machioneetal2018}
{\sc Machicoane, N., Labarre, V., Voisin, B., Moisy, F. \& Cortet, P.-P.} 2018
  Wake of inertial waves of a horizontal cylinder in horizontal translation.
  {\em Phys. Rev. Fluids\/} {\bf 3}, 034801.

\bibitem[de~Marez {\em et~al.\/}(2017)de~Marez, Carton, Morvan \&
  Reinaud]{DeMarezetal17}
{\sc de~Marez, C., Carton, X., Morvan, M. \& Reinaud, J.} 2017 The interaction
  of two surface vortices near a topographic slope in a stratified ocean. {\em
  Fluids\/} {\bf 2}~(57).

\bibitem[de~Marez {\em et~al.\/}(2020)de~Marez, Morvan, L'H{\'{e}}garet,
  Meunier \& Carton]{DeMarezetal20}
{\sc de~Marez, C., Morvan, M., L'H{\'{e}}garet, P., Meunier, T. \& Carton, X.}
  2020 Vortex-wall interaction on the beta-plane and the generation of deep
  submesoscale cyclones by internal {K}elvin waves-current interactions. {\em
  Geophys. Astrophys. Fluid Dyn.\/} {\bf 114}~(4-5), 588--606.

\bibitem[Meleshko \& van Heijst(1994)]{MeleshkoH94}
{\sc Meleshko, V.~V. \& van Heijst, G. J.~F.} 1994 On {C}haplygin's
  investigations of two-dimensional vortex structures in an inviscid fluid.
  {\em J. Fluid Mech.\/} {\bf 272}, 157--182.

\bibitem[Moffatt(1969)]{MOFFATT69}
{\sc Moffatt, H.~K.} 1969 The degree of knottedness of tangled vortex lines.
  {\em J. Fluid Mech.\/} {\bf 35}~(1), 117--129.

\bibitem[Penduff {\em et~al.\/}(2011)Penduff, Juza, Barnier, Zika, Dewar,
  Treguier, Molines \& Audiffren]{PENDUFFETAT}
{\sc Penduff, T., Juza, M., Barnier, B., Zika, J., Dewar, W.~K., Treguier,
  A.-M., Molines, J.-M. \& Audiffren, N.} 2011 Sea level expression of
  intrinsic and forced ocean variabilities at interannual time scales. {\em J.
  Climate\/} {\bf 24}~(21), 5652--5670.

\end{thebibliography}

\end{document}